\documentclass[a4paper,12pt]{article}
\pdfoutput=1
\usepackage{amsfonts,amsthm,amsmath,amssymb,graphicx}
\usepackage{color}
\usepackage[T1]{fontenc}
\usepackage{lmodern}
\usepackage[utf8]{inputenc}

\usepackage[unicode]{hyperref}
\hypersetup{
    hypertexnames=false,
    colorlinks=false,
    unicode=true,
    pdfborder = {0 0 0},
    pdfstartview={FitH},
    pdfauthor={Pawe{\l} Caputa, Joan Sim\'on, Andrius \v{S}tikonas, Tadashi Takayanagi and Kento Watanabe},
    pdftitle={Scrambling time from local perturbations of the eternal BTZ black hole},
    bookmarksnumbered=true,
    linktoc=all
}

\begin{document}

\newtheorem{definition}{Definition}[section]
\newcommand{\be}{\begin{equation}}
\newcommand{\ee}{\end{equation}}
\newcommand{\bea}{\begin{eqnarray}}
\newcommand{\eea}{\end{eqnarray}}
\newcommand{\LE}{\left[}
\newcommand{\R}{\right]}
\newcommand{\nn}{\nonumber}
\newcommand{\Tr}{\text{Tr}}
\newcommand{\N}{\mathcal{N}}
\newcommand{\vf}{\varphi}
\newcommand{\LL}{\mathcal{L}}
\newcommand{\Op}{\mathcal{O}}
\newcommand{\HH}{\mathcal{H}}
\newcommand{\arctanh}{\text{arctanh}}
\newcommand{\up}{\uparrow}
\newcommand{\down}{\downarrow}
\newcommand{\ket}[1]{\left| #1 \right>}
\newcommand{\bra}[1]{\left< #1 \right|}
\newcommand{\ketbra}[1]{\left|#1\right>\left<#1\right|}
\newcommand{\rd}{\partial}
\newcommand{\de}{\partial}
\newcommand{\ba}{\begin{eqnarray}}
\newcommand{\ea}{\end{eqnarray}}
\newcommand{\db}{\bar{\partial}}
\newcommand{\we}{\wedge}
\newcommand{\ca}{\mathcal}
\newcommand{\lr}{\leftrightarrow}
\newcommand{\f}{\frac}
\newcommand{\s}{\sqrt}
\newcommand{\vp}{\varphi}
\newcommand{\hvp}{\hat{\varphi}}
\newcommand{\tvp}{\tilde{\varphi}}
\newcommand{\tp}{\tilde{\phi}}
\newcommand{\ti}{\tilde}
\newcommand{\ap}{\alpha}
\newcommand{\pr}{\propto}
\newcommand{\mb}{\mathbf}
\newcommand{\ddd}{\cdot\cdot\cdot}
\newcommand{\no}{\nonumber \\}
\newcommand{\la}{\langle}
\newcommand{\lb}{\rangle}
\newcommand{\ep}{\epsilon}
 \def\we{\wedge}
 \def\lr{\leftrightarrow}
 \def\f {\frac}
 \def\ti{\tilde}
 \def\ap{\alpha}
 \def\pr{\propto}
 \def\mb{\mathbf}
 \def\ddd{\cdot\cdot\cdot}
 \def\no{\nonumber \\}
 \def\la{\langle}
 \def\lb{\rangle}
 \def\ep{\epsilon}

\begin{titlepage}
\thispagestyle{empty}

\begin{flushright}
NORDITA-2015-32\\
YITP-15-21\\
IPMU15-0033\\
\end{flushright}

\vspace{.4cm}
\begin{center}
\noindent{\Large \textbf{Scrambling time from local perturbations}}\\
\vspace{0.5cm}
\noindent{\Large \textbf{of the eternal BTZ black hole}}
\vspace{2cm}

Pawe{\l} Caputa$^{a}$, Joan Sim\'on$^{b}$, Andrius \v{S}tikonas$^{b}$, \\
Tadashi Takayanagi$^{c,d}$ and Kento Watanabe$^{c}$
\vspace{1cm}

{\it
$^{a}$Nordita, KTH Royal Institute of Technology and Stockholm University,\\
Roslagstullsbacken 23, SE-106 91 Stockholm, Sweden\\
$^{b}$ School of Mathematics and Maxwell Institute for Mathematical Sciences,\\
University of Edinburgh, King's Buildings,
Edinburgh EH9 3FD, UK\\
$^{c}$Yukawa Institute for Theoretical Physics (YITP),
Kyoto University, Kyoto 606-8502,\\
$^{d}$Kavli Institute for the Physics and Mathematics of the Universe (Kavli IPMU),\\
University of Tokyo, Kashiwa, Chiba 277-8582, Japan\\
}

\vskip 2em
\end{center}

\vspace{.5cm}
\begin{abstract}
We compute the mutual information between finite intervals in two non-compact 2d CFTs in the thermofield double formulation after one of them has been locally perturbed by a primary operator at some time $t_\omega$ in the large $c$ limit. We determine the time scale, called the scrambling time, at which the mutual information vanishes and the original entanglement between the thermofield double gets destroyed by the perturbation.  We provide a holographic description in terms of a free falling particle in the eternal BTZ black hole that exactly matches our CFT calculations. Our results hold for any time $t_\omega$. In particular, when the latter is large, they reproduce the bulk shock-wave propagation along the BTZ horizon description.
\end{abstract}

\end{titlepage}

\tableofcontents

\newpage

\section{Introduction and Summary}

Consider a (1+1)d CFT in some state having non-trivial quantum correlations. If the system is perturbed at some instant of time $t_\omega$ and evolve unitarily afterwards, it is natural to ask whether there exists any time scale when its subsystems become uncorrelated. Since the mutual information $I_{A:B}=S_{A}+S_{B}-S_{A\cup B}$ between two such subsystems $A$ and $B$ provides an upper bound for the connected two-point functions of operators acting on these subsystems\footnote{This bound is proved for finite dimensional Hilbert spaces. We are not aware of an extension of this result to QFTs/CFTs, but we expect it to hold when regulating and normalising appropriately the relevant quantities in the continuum limit.} \cite{cirac}
\begin{equation}
  I_{A:B} \geq \frac{\left(\langle {\mathcal O}_A{\mathcal O}_B\rangle - \langle {\mathcal O}_A\rangle \langle {\mathcal O}_B\rangle\right)^2}{2\|{\mathcal O}_A\|^2\|{\mathcal O}_B\|^2}\,,
\label{eq:bound}
\end{equation}
it is natural to study the vanishing of this quantity to answer this question. The study of the time dependence in this measure of entanglement can help us to understand the time scales controlling how quantum systems get thermalized, which is one of the most important problems in non-equilibrium physics.

A particular situation of the above scenario is when a perturbation acts on a thermal state. In holographic theories, thermal states are believed to have a gravity dual in terms of black holes \cite{Witten:1998zw,eternal}. Black hole physics suggests that the speed at which the system forgets initial conditions, i.e. the perturbation, is the fastest among all physical diffusive processes. This gave rise to the notion of fast scramblers and the scrambling conjecture \cite{Hayden:2007cs,Sekino:2008he}. The main goal of this paper is to provide a first principle derivation for the time scale at which this phenomenon occurs for 2d CFTs in the large $c$ limit in a concrete setup which allows both CFT and holographic computations.

Recently, Shenker and Stanford considered an excellent and tractable setup to study the fast scrambling phenomena in the context of an eternal black hole \cite{Shenker:2013pqa,Shenker:2013yza,Shenker:2014cwa}. This involves a pair of non-interacting CFTs in an entangled state, the thermofield double state. Tracing any entire CFT Hilbert space, gives rise to a thermal density matrix in the remaining CFT. The perturbation is described by some boundary CFT operator and its gravity dual involved a shock-wave propagating in the black hole background. No matter how small the boundary perturbation is, the blue shift of energies when this perturbation reaches the horizon suggests the existence of a non-trivial backreaction.

In this work, we study such a setup for a perturbation localized in a point-like region, triggered by a primary operator in a given CFT. To obtain analytical results, we consider 2d large $c$ CFTs and their gravity duals given by a perturbation of the BTZ black hole \cite{BTZ}. Recent developments in the calculation of 4-pt functions involving heavy and light operators in the large $c$ limit of the dual 2d CFT \cite{Fitzpatrick:2014vua,Fitzpatrick:2015zha} (see also \cite{Hartman:2013mia,Hijano:2015rla}) allow us to analytically test these ideas.

Computations of time evolutions of entanglement entropy after local perturbations\footnote{Note that this setup looks similar to the local quenches in CFTs
\cite{cal}. However, in the latter the local excitations are triggered by joining
two semi-infinite lines and lead to local excitations in all sectors of a given CFT.
Thus their behaviours differ from each other in integrable CFTs \cite{Nozaki:2014hna,He:2014mwa,Nozaki:2014uaa}. On the other hand, in large $c$ CFTs, they behave similarly \cite{NNT,Caputa:2014vaa,Asplund:2014coa} in that both results for 2d CFTs show logarithmic time evolution of entanglement entropy.} by primary operators have been formulated in \cite{Nozaki:2014hna,Nozaki:2014uaa} and have been applied to many examples for CFTs at zero temperature in \cite{He:2014mwa,Caputa:2014vaa,Asplund:2014coa,Giusto:2014aba,deBoer:2014sna,Guo:2015uwa}.
Entanglement entropy and mutual information at finite temperature CFTs has been analyzed for integral CFTs in \cite{Caputa:2014eta}. On the other hand, the holographic calculations of time-evolutions of entanglement entropy after local perturbations have been analyzed in \cite{NNT,Ugajin:2013xxa,Asplund:2013zba} at zero temperature and in \cite{Caputa:2014eta} at finite temperature. In this paper we will extend the discussion of local excitations to the thermofield double formalism of finite temperature CFTs in the large $c$ limit.

\paragraph{Summary of results :} Specifically, we perturb the thermofield double (TFD) state by a local primary operator $\psi$ at time $t_\omega$ in the past and compute the mutual information between regions $A$ and $B$ belonging to opposite boundaries. We denote the two boundary times in the thermofield double by $t_L\equiv t_-$ and $t_R\equiv t_+$. When measuring the mutual information at $t_-=t_+=0$, we ask for the time scale $t^\star_\omega$ when the mutual information vanishes
\begin{equation}
  I_{A:B}(t^\star_\omega)=0\,.
\end{equation}
Equivalently, we ask for the time scale $t^\star_\omega$ at which correlations between $A$ and $B$ vanish. When both subsystems $A$ and $B$ are the intervals $(0<)y\leq x\leq y+L$ and the perturbation is turned on at $x=0$ and time $t_-=-t_\omega$, we obtain the following analytical result for $t^\star_\omega\gg\beta$
\begin{equation}
t^\star_\omega=y+\frac{L}{2}-\frac{\beta}{2\pi}\log\left(\frac{\beta}{\pi\epsilon}\frac{\sin\pi \alpha_\psi}{\alpha_\psi}\right)+\frac{\beta}{\pi}\log\left(2 \sinh\frac{\pi L}{\beta}\right), \label{scrq}
\end{equation}
where $\beta$ is the inverse temperature and $\alpha_{\psi} = \sqrt{1-24h_\psi/c}$ carries the information about the primary operator perturbation of conformal dimension $h_\psi$. The parameter $\ep$ represents a UV cut off for the local excitation, so that the excited state is localized around a region of size $\epsilon$ of the operator insertion. This makes the energy of the perturbation $E_\psi = \f{\pi h_{\Psi}}{\ep}$ finite. In the limit $h_\psi/c\ll 1$, which is the relevant one to match the butterfly effect discussed in \cite{Shenker:2013pqa,Shenker:2013yza}, this reduces to
\begin{equation}
  t^\star_w = f(L,\beta) + \frac{\beta}{2\pi}\log \left(\frac{\pi S_{density}}{4 E_\psi}
  \right),  \label{scrqg}
\end{equation}
where $S_{density}=\f{\pi c}{3\beta}$ is the entropy density of the original thermal system. The $\log S$ behavior in \eqref{scrqg} is consistent with the fast scrambling conjecture \cite{Sekino:2008he,Shenker:2013pqa}.

Given the bound \eqref{eq:bound}, it should be possible to extract the same time scale from the condition of vanishing two sided 2-pt functions. We explicitly show this in appendix \ref{ap-C}, confirming the observation made in Shenker and Stanford \cite{Shenker:2013pqa} that both scales are controlled by the same physics.

In the second part of this work, we derive the same time scale from bulk holographic considerations and find a {\it perfect matching} between both calculations. Our holographic model is based on the description of the local boundary perturbation in terms of some free falling particle satisfying an initial condition guaranteeing such particle carries the right amount of energy from the CFT stress tensor perspective. This is done by generalizing the model in \cite{NNT,Caputa:2014eta} to the two sided BTZ black hole, based on the back reaction description of point particles as quotients of AdS${}_3$. Applying the holographic entanglement entropy \cite{RTorig,Hubeny:2007xt} to evaluate the entanglement entropy and mutual information in our set-up, leads to the same scrambling time \eqref{scrq}. At the same time, our setup and calculation may be useful for the interesting question regarding the dual CFT interpretation of a particle falling into a AdS black hole in future studies. 

Our local boundary perturbation includes a regularization parameter $\ep$ describing its size. This parameter is holographically interpreted as the bulk position (distance from the boundary) from which the massive particle falls into the black hole. Our solution computes the back reacted geometry for any $t_\omega$ and approaches a localised shock-wave in the limit of large $t_\omega$ \cite{Roberts:2014isa,Shenker:2013pqa}.

The paper is organized as follows: In section 2, we will analyze the time evolution of entanglement entropy in large $c$ 2d CFTs at finite temperature, which agrees perfectly with a previous gravity dual computation. In section 3, we study local perturbations in finite temperature CFTs by employing the thermofield double formalism. We compute the mutual information from entanglement entropies. In section 4, we compute the scrambling time for the mutual information. In section 5, we introduce our holographic model. In section 6,  we present our holographic computations of mutual information in a two sided  $AdS_3$ black hole background with a local excitation. In appendix A, we explained the details of treatment of twist operators in the replica method computations of entanglement entropy in the thermofield double formulation. In appendix B, we present some details of our holographic model. In appendix C we describe a computation of two point function in our model.

\smallskip
{\bf Note Added:} While finishing our main computations, the work of Roberts and Stanford \cite{Roberts:2014ifa} appeared. The latter has a detailed account of two point functions in the presence of localised excitations over thermal states and briefly mentions the behaviour of the mutual information in the same set-up. Thus, it has some overlap with our results.
 In our paper, we literally evaluate the mutual information between the thermofield double in both 2d large $c$ CFTs and their gravity duals independently and show their results perfectly agree. Our gravity solutions explicitly have the regularization parameter $\ep$ and our matching between gravity and CFT results holds while keeping this parameter small but non-zero. We would also like to mention that in the interesting recent paper \cite{Maldacena:2015waa} by Maldacena, Shenker and Stanford, the fast scrambling behavior of the correlations functions has been interpreted in terms of chaos.

\section{Single sided entropy}

To introduce our basic tools and fix the notation, we analyse the local perturbation to a thermal state in a single 2d CFT at finite temperature in this section.\\
Consider a thermal state $\rho_\beta$ locally perturbed by a primary operator $\psi(0,-t_\omega)$ inserted at $x=0$ at time $-t_\omega$. The time evolution of the resulting density matrix is given by
\begin{equation}
  \rho(t) = {\cal N}\,e^{-iHt}\psi(0,-t_\omega)\,\rho_\beta\,\psi^\dagger(0,-t_\omega)\,e^{iHt}\,.
\end{equation}
where $H$ is the Hamiltonian of our system.\\
Denote by $\rho_A = \text{Tr}_{\bar{A}} \rho(t)$ the reduced density matrix on a finite interval $A$ with endpoints\footnote{Notice that the perturbation is originally inserted outside of the interval $A$.} $y,\,y+L$ satisfying $y,L > 0$. Its entanglement entropy $S_A$ can be computed using the replica trick. We first compute the Renyi entropies
\begin{equation}
S^{(n)}_A\equiv \frac{1}{1-n}\log\Tr \rho^n_A(t)\,.
\label{RenyiE}
\end{equation}
The entanglement entropy is obtained by taking the limit $S_A=\lim_{n\to1} S^{(n)}_A$.\\
The trace of the reduced density matrix $\Tr\rho^n_A(t)$ requires the calculation of the normalised 4-point function
\begin{equation}
\Tr\rho^n_A(t)=\frac{\langle\Psi(x_1,\bar{x}_1)\sigma_n(x_2,\bar{x}_2)\tilde{\sigma}_n(x_3,\bar{x}_3)\Psi^{\dagger}(x_4,\bar{x}_4)\rangle}{\left(\langle\psi(x_1,\bar{x}_1)\psi^{\dagger}(x_4,\bar{x}_4)\rangle_{C_1}\right)^n}
\label{eq:r1}
\end{equation}
with the insertion points
\begin{eqnarray}
x_1=-i\epsilon,\quad x_2=y-t_\omega-t_-,\quad x_3=y+L-t_\omega-t_-,\quad x_4=+i\epsilon\nn\\
\bar{x}_1=+i\epsilon,\quad \bar{x}_2=y+t_\omega+t_-,\quad \bar{x}_3=y + L+t_\omega+t_-,\quad \bar{x}_4=-i\epsilon\,.
\label{eq:shift1}
\end{eqnarray}
The operator $\Psi$ stands for the product of the operators $\psi_i$ in each of the i-th copies of the theory\footnote{This correlator is formally computed in the cyclic orbifold $CFT^n/Z_n$.}
\begin{equation}
\Psi = \psi_1 \cdot \psi_2 \cdots \psi_n\label{PSI}
\end{equation}
and has conformal dimension $h_\Psi = n\,h_\psi$, where $h_\psi$ is the conformal dimension of the original perturbation $\psi$. Notice $\epsilon$ is a parameter smearing the local operator perturbation and all the time evolution is carried by the twist operators $\sigma_n,\,\tilde{\sigma}_n$ which are initially inserted at both ends of the interval when cyclically gluing the different cylinder copies that give rise to the manifold $C_n$. Finally, the conformal dimension $\Delta_\sigma=2H_\sigma$ of the twist operators is
\begin{equation}
H_\sigma=\frac{c}{24}\left(n-\frac{1}{n}\right)\,.
\end{equation}

\medskip
We compute \eqref{eq:r1} analogously to \cite{Asplund:2014coa} but with an additional composition of a map from the cylinder to the plane
\begin{equation}
w(x)=e^{\frac{2\pi}{\beta}x},
\label{eq:c-plane}
\end{equation}
to take care of the thermal nature of the original state, as well as the map
\begin{equation}
z(w)=\frac{(w_1-w)w_{34}}{w_{13}(w-w_4)}
\end{equation}
that brings the points $w_1\to 0$, $w_2\to z$, $w_3\to1$, $w_4\to \infty$.\\
The transformation properties of primary operators determine the resulting trace to be\footnote{We already used the regularised twist operators so that $\varepsilon_{UV}$ is the standard UV cut-off.}

\begin{equation}
\Tr\rho_A^n(t)=\left|\frac{\beta}{\pi \varepsilon_{UV}}\sinh\left(\frac{\pi x_{23}}{\beta}\right)\right|^{-4H_\sigma}\left|1-z\right|^{4H_\sigma}G(z,\bar{z})
\label{eq:one}
\end{equation}
where we used the 2-pt function on the cylinder $C_1$
\begin{equation}
\langle \psi(x_1,\bar{x}_1)\psi(x_4,\bar{x}_4)\rangle_{C_1}=\left|\frac{\beta}{\pi}\sinh\left(\frac{\pi x_{14}}{\beta}\right)\right|^{-4h_\psi}
\end{equation}
and introduced the canonical 4-point function
\begin{equation}
\begin{aligned}
G(z,\bar{z})&\equiv\lim_{z_4\to \infty}|z_4|^{4h_\Psi}\langle\psi(z_4,\bar{z}_4)\sigma_n(z,\bar{z})\tilde{\sigma}_n(1,1)\psi(0,0)\rangle \\
&\equiv \langle\psi|\sigma_n(z,\bar{z})\tilde{\sigma}_n(1,1)|\psi\rangle
\label{eq:mcor}
\end{aligned}
\end{equation}
defined in terms of the cross-ratio
\begin{equation}
z=\frac{w_{12}w_{34}}{w_{13}w_{24}},\qquad 1-z=\frac{w_{14}w_{23}}{w_{13}w_{24}}\,.
\end{equation}
We defined $w_{ij}=w_i-w_j$ in all the above formulas and the same conventions hold for $x_{ij}$ and $z_{ij}$.\\
The Renyi entropies are computed by inserting \eqref{eq:one} into \eqref{RenyiE}
\begin{equation}
S^{(n)}_A=\frac{c(n+1)}{6}\log\left(\frac{\beta}{\pi\varepsilon_{UV}}\sinh\frac{\pi\,L}{\beta}\right)+\frac{1}{n-1}\log(|1-z|^{4H_\sigma}G(z,\bar{z}))\, ,
\end{equation}
where $\varepsilon_{UV}$ is the UV cut off of the CFT i.e. the lattice spacing. The first term is the standard Renyi entropy for an interval $L$ in a 2d CFT at finite temperature $T=1/\beta$; the second term captures the extra contribution due to the local operator insertion. In particular, the dependence on the conformal dimension of the local operator $h_\psi$ is encoded in $G(z,\bar{z})$.

In general, the extra contribution to the Renyi entropies requires the knowledge of the full four-point function $G(z,\bar{z})$, i.e. the dynamical details of the particular 2d CFT under consideration. To make further progress, we consider the large $c$ limit\footnote{We assume a class of CFTs allowing such a limit.}. Notice that in the limit $n\to 1$, the twist operators $\sigma_n,\,\tilde{\sigma}_n$ become light
\begin{equation}
  H_\sigma/c=\frac{1}{24}\left(n-\frac{1}{n}\right) \to 0 \quad \text{as} \quad n\to 1
\end{equation}
If $h_\psi/c$ remains fixed in the large $c$ limit, the 4-pt function \eqref{eq:mcor} becomes a 4-pt function involving two heavy and two light operators. This is precisely the set-up considered in \cite{Fitzpatrick:2014vua,Fitzpatrick:2015zha} to compute the dominant (vacuum) contribution to $G(z,\bar{z})$\footnote{We refer the readers to \cite{CC,Cardy:2007mb} for the description of the regularisation of twist operators used to compute entanglement entropy.}. Using their results, one derives \cite{Asplund:2014coa}
\begin{equation}
\log G(z,\bar{z})\simeq -\frac{c(n-1)}{6}\log\left(\frac{z^{\frac{1}{2}(1-\alpha_\psi)}\bar{z}^{\frac{1}{2}(1-\bar{\alpha}_\psi)}(1-z^{\alpha_\psi})(1-\bar{z}^{\bar{\alpha}_\psi})}{\alpha_\psi\bar{\alpha}_\psi}\right)+{\cal O}((n-1)^2)
\end{equation}
where
\begin{equation}
\alpha_\psi=\sqrt{1-\frac{24h_\psi}{c}}\,,
\end{equation}
encodes all the dependence on the conformal dimension of the local operator $h_\psi$. Finally, we can compute the variation in the entanglement entropy due to the insertion of the local primary operator to be
\begin{equation}
\Delta S_A=\frac{c}{6}\log\left(\frac{z^{\frac{1}{2}(1-\alpha_\psi)}\bar{z}^{\frac{1}{2}(1-\bar{\alpha}_\psi)}(1-z^{\alpha_\psi})(1-\bar{z}^{\bar{\alpha}_\psi})}{\alpha_\psi\bar{\alpha}_\psi(1-z)(1-\bar{z})}\right)\,,
\label{eq:dsa}
\end{equation}
where we subtracted the entanglement entropy of the interval $L$ at finite temperature $T=1/\beta$
\begin{equation}
S_{\text{thermal}}=\frac{c}{3}\log\left(\frac{\beta}{\pi\varepsilon_{UV}}\sinh\frac{\pi\,L}{\beta}\right)\,.
\label{eq:thermal}
\end{equation}

As explained in \cite{Caputa:2014eta}, to extract a non-trivial contribution to the entanglement entropy in the CFT at finite temperature we must take the smearing parameter $\epsilon$ small but finite. This still allows us to work with completely analytic formulas at order $\epsilon$. Then, in the small $\epsilon/\beta$ limit\footnote{We are assuming that $y-t_- -t_\omega$ and $y+L-t_- -t_\omega$ are larger than the smearing parameter $\epsilon$ in units of $\beta$.}, the cross-ratios are
\begin{equation}
\begin{aligned}
  z&\simeq 1+\frac{2\pi i \epsilon}{\beta}\frac{\sinh\frac{\pi\,L}{\beta}}{\sinh\frac{\pi (y+L-t_- -t_\omega)}{\beta}\sinh\frac{\pi (y-t_- -t_\omega)}{\beta}}+{\cal O}(\epsilon^2) \\
  \bar{z}& \simeq 1-\frac{2\pi i \epsilon}{\beta}\frac{\sinh\frac{\pi\,L}{\beta}}{\sinh\frac{\pi (y+L+t_- +t_\omega)}{\beta}\sinh\frac{\pi (y+t_- +t_\omega)}{\beta}}+{\cal O} (\epsilon^2)
\label{eq:zA}
\end{aligned}
\end{equation}
Due to the non-trivial monodromy properties of $G (z, \bar{z})$, we must carefully deal with the sign of the imaginary part of the cross-ratios \cite{Asplund:2014coa,Roberts:2014ifa}\footnote{We choose a reference phase to be consistent with causality and make entanglement entropies real and non-negative.}. Notice the imaginary part of $\bar{z}$ never changes sign, for $t_-+t_\omega \geq 0$. Thus, we conclude $\bar{z}\simeq 1$ for all such times. On the other hand, the imaginary part of $z$ does flip sign whenever $t_-+t_\omega \in (y,\,y+L)$. Thus, we either have $(z,\bar{z})\to (1,1)$ for $t_-+t_\omega<y$ and $t_-+t_\omega>y+L$ or  $(z,\bar{z})\to (e^{2\pi i},1)$ for $y < t+t_\omega < y+ L$. Using these phases in \eqref{eq:dsa}, we reach our first important result
\begin{equation}
\boxed{
\begin{aligned}
\Delta S_A &=0\,, \quad t_-+t_\omega < y \,\,\text{and}\,\, t_-+t_\omega > y + L \\
\Delta S_A &=\frac{c}{6}\log\left[\frac{\beta}{\pi\epsilon}\frac{\sin \pi \alpha_\psi}{\alpha_\psi}\frac{\sinh\left(\frac{\pi(y+L -t_- -t_\omega)}{\beta}\right)\sinh\left(\frac{\pi(t_- +t_\omega -y)}{\beta}\right)}{\sinh\left(\frac{\pi L}{\beta}\right)}\right] \quad \quad y<t_- +t_\omega < y+L\,.
\end{aligned}
}\label{eq:S_A_CFT}
\end{equation}
Thus, there is no variation in the entanglement entropy $S_A$ either till the perturbation reaches region $A$ $(t_-+t_\omega < y)$ or when it leaves region $A$ $(t_-+t_\omega > y + L)$. While the perturbation can causally be in region $A$, the variation in entanglement reaches a maximum at $t_-+t_\omega = y + \frac{L}{2}$
\begin{equation}
  (\Delta S_A)_\text{max} = \frac{c}{6}\log\left[\frac{\beta}{2\pi\epsilon}\frac{\sin \pi \alpha_\psi}{\alpha_\psi}\tanh\frac{\pi\,L}{2\beta}\right]\,.
\end{equation}
In the high temperature limit (or large interval $L$), the increase in entanglement due to the perturbation equals
\begin{equation}
  \Delta S_A \simeq \frac{c}{6}\log\left[\frac{\beta}{2\pi\epsilon}\frac{\sin\pi\alpha_\psi}{\alpha_\psi}\right] \qquad \beta \to 0\,,\,\,\frac{\beta}{\epsilon} \gg 1\,.
\end{equation}

In section \ref{SAGr} we will match this result with the gravity dual computation using the holographic entanglement entropy. Let us now proceed with the entanglement entropies that involve intervals in both CFTs.

\section{Two-sided entropies}
Entanglement entropy and mutual information in the thermofield double state involving intervals in both CFTs were discussed in detail in \cite{Mor} and their time evolution in \cite{HaMa,Caputa:2013eka}.  Here we only briefly review this setup and extend by the insertion of a local operator to one of the CFTs.

Consider two non-interacting 2d CFTs, $CFT_L$ and $CFT_R$
with isomorphic Hilbert spaces $\mathcal{H}_L$ and $\mathcal{H}_R$.
The thermofield double (TFD) state is a particular entangled state in the total Hilbert space
$\mathcal{H}_{tot} = \mathcal{H}_L \otimes \mathcal{H}_R$.
\begin{equation}
| \Psi_\beta \rangle
= \frac{1}{\sqrt{Z (\beta)}} \sum_n e^{-\frac{\beta}{2} E_n} |n \rangle_L |n \rangle_R ,
\end{equation}
where $| n \rangle_{L, R} \in \mathcal{H}_{L, R}$ are the eigenstates in each Hilbert space
and $Z (\beta)$ is a partition function in total Hilbert space (and also in each Hilbert space).
\begin{equation}
Z (\beta) = \sum_n e^{-\beta E_n}
\end{equation}
If we have the total Hamiltonian $H_{tot} = H_L + H_R$,
this partition function is a partition function on a cylinder $C_1$ with circumference $\beta$
\begin{equation}
Z (\beta)
= \sum_{n, m} \langle n |_L \langle n |_R e^{-\frac{\beta}{2} (H_L + H_R) } |m \rangle_L |m \rangle_R
= \mathrm{Tr}_{tot} \left[ e^{-\frac{\beta}{2} (H_L + H_R)} \right]
\end{equation}
Tracing out the Hilbert space $\mathcal{H}_R$ from the pure state density matrix,
we can get the thermal density matrix in $CFT_L$ with temperature $\beta$
\begin{equation}
\rho_L
= \mathrm{Tr}_R | \Psi_\beta \rangle \langle \Psi_\beta |
= \frac{1}{Z (\beta)} \sum_n e^{-\beta E_n} |n \rangle_L \langle n |_L
\end{equation}
A general time evolution of the TFD state is obtained by applying the evolution operator to both CFTs
\begin{equation}
| \Psi_\beta (t_-, t_+)\rangle
= e^{-i t_- H_L + i t_+ H_R} | \Psi_\beta \rangle
= \frac{1}{\sqrt{Z (\beta)}} \sum_n e^{-i(t_+ - t_- -i \frac{\beta}{2}) E_n} |n \rangle_L |n \rangle_R
\end{equation}
One can immediately check that setting $t_-=t_+=t$, that can be seen as evolving with Hamiltonian $H_L-H_R$, leaves the TFD state invariant (the symmetry of the TFD state). On the other hand, setting $t_-=-t_+=t$ yields the time dependent state corresponding to evolution with $H_L+H_R$ as in \cite{HaMa}. These two configurations should be kept in mind since we leave general $t_-$ and $t_+$ in our formulas so that our formalism can be used to extract the evolution after a local excitation with any of the two total Hamiltonians\footnote{In fact our results also hold for evolution of the TFD state with $H_L$ or $H_R$ only and one can extract these formulas by setting $t_-=t$ and $t_+=0$ or $t_-=0$ and $t_+=t$ respectively.}.

\begin{figure}[ht]
 \begin{minipage}{0.5\hsize}
  \begin{center}
   \includegraphics[natwidth=85, width=85mm]{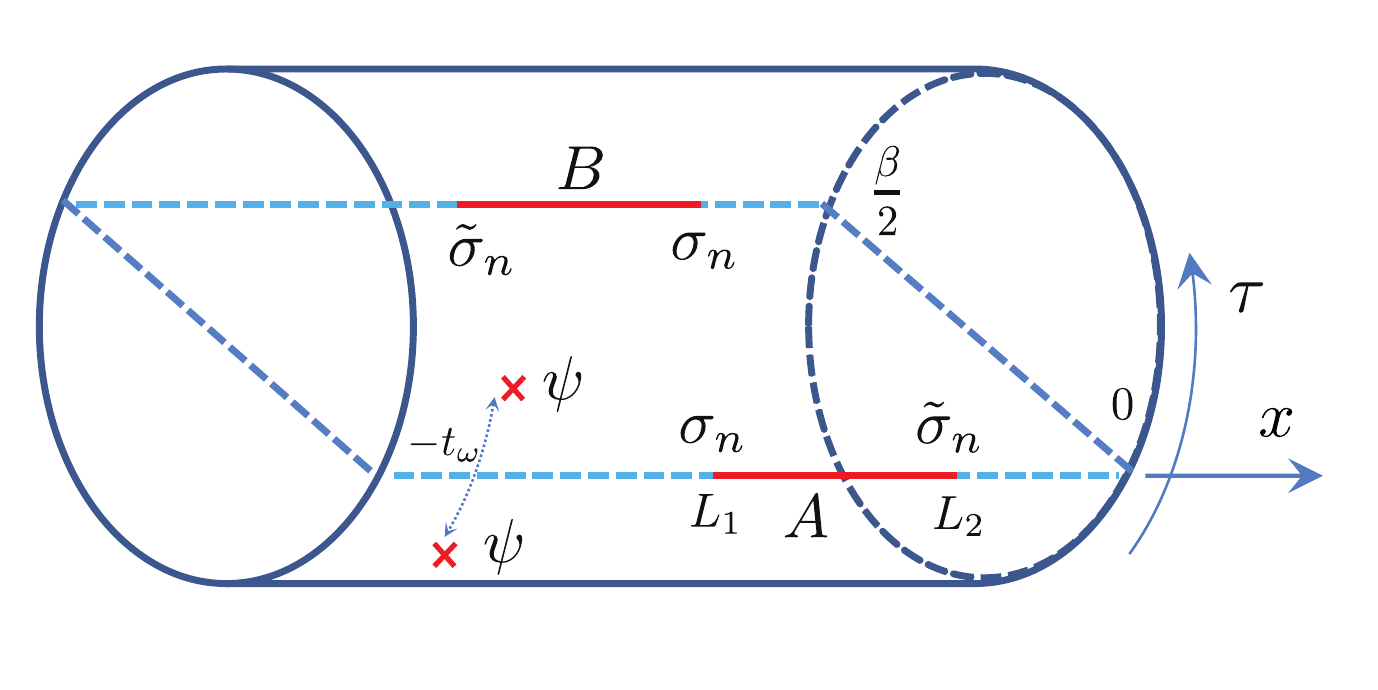}
  \end{center}
 \end{minipage}
  \begin{minipage}{0.40\hsize}
  \begin{center}
   \includegraphics[natwidth=65mm, width=65mm]{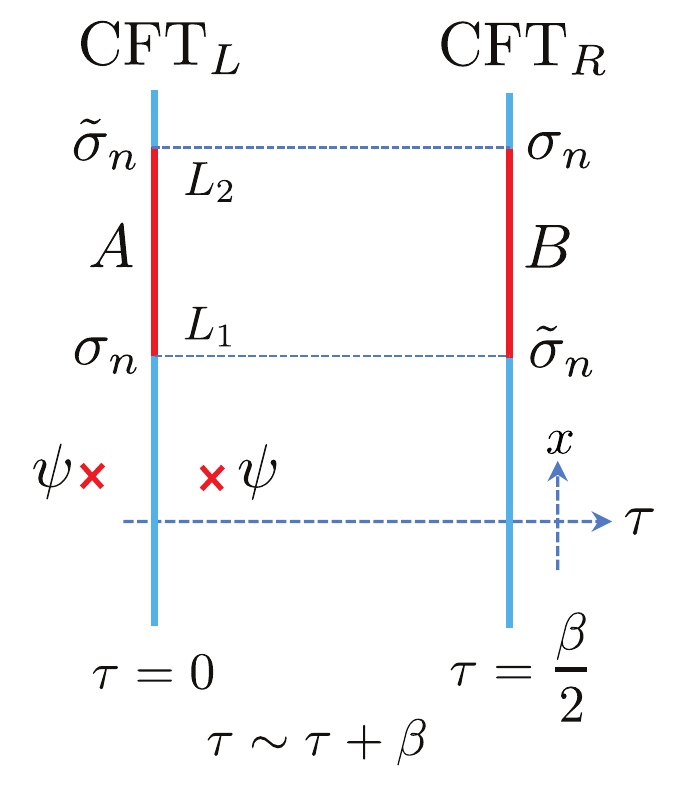}
  \end{center}
 \end{minipage}
 \caption{Our setup in the computation of the mutual information. We have two intervals $A$ and $B$ of size $L_2-L_1=L$ in each CFT and a local operator inserted at time $t_w$ in the past. The operators are separated by distance $2\epsilon$ and in the CFT formulas we use $L_1=y$ and $L_2=y+L$.}\label{twoside}
\end{figure}

In the TFD formalism, we can also relate one-sided and two-sided correlators by the analytical continuation of $t$.
For example, consider the following one-sided correlator
\begin{equation}
\langle \Psi_{\beta} | \mathcal{O}_L (x_1, 0) \mathcal{O}_L^{\dagger} (x_2, t)  | \Psi_{\beta} \rangle
= \sum_{n, m} e^{-\beta E_n + i t (E_n - E_m)}
\langle n |_L \mathcal{O}_L (x_1, 0) | m \rangle_L
\langle  m|_L \mathcal{O}_L^{\dagger} (x_2, 0) | n \rangle_L
\end{equation}
On the other hand, we can deform the two-sided correlator as follows
\begin{align}
\langle \Psi_{\beta} | \mathcal{O}_L (x_1, 0) \mathcal{O}_R (x_2, t)  | \Psi_{\beta} \rangle
&= \sum_{n, m} e^{-\frac{\beta}{2} (E_n + E_m) + i t (E_n - E_m)}
\langle n |_L \mathcal{O}_L (x_1, 0) | m \rangle_L
\langle n |_R \mathcal{O}_R (x_2, 0) | m \rangle_R \nn \\
&= \sum_{n, m} e^{-\beta E_n + i (t  - i \frac{\beta}{2}) (E_n - E_m)}
\langle n |_L \mathcal{O}_L (x_1, 0) | m \rangle_L
\langle m |_R \mathcal{O}_R^{\dagger} (x_2, 0) | n \rangle_R
\end{align}
Therefore, the one-sided and two-sided correlators can be related
through the analytical continuation $t \to t + i \frac{\beta}{2}$
\begin{equation}
\langle \Psi_{\beta} | \mathcal{O}_L (x_1, 0) \mathcal{O}_L (x_2, t)  | \Psi_{\beta} \rangle
= \langle \Psi_{\beta} | \mathcal{O}_L (x_1, 0) \mathcal{O}_R^{\dagger} (x_2, t - i \frac{\beta}{2}) | \Psi_{\beta} \rangle  \label{eq2s}
\end{equation}
If the operators $\mathcal{O}_L$ in $CFT_L$ are located at $\tau = 0$,
the operators $\mathcal{O}_R$ are located at $\tau = \frac{\beta}{2}$,
or at the opposite side on the cylinder.
We can express the correlators in the TFD state as the correlators on the cylinder $C_1$ (see Fig. \ref{twoside}).

Let us also mention a simple fact related to the symmetry of the TFD state. Namely, in the CFT, we compute the entanglement entropies as well as the mutual information in a state
\begin{equation}
\tilde{\ket{\psi}}=e^{-iH_Lt_\omega}\mathcal{O}_L(x)e^{iH_Lt_\omega}\ket{\psi_\beta}
\end{equation}
with the TFD state $\ket{\psi_\beta}$. Since $H_L-H_R$ leaves the TFD invariant, the above state is equivalent to
\begin{equation}
\tilde{\ket{\psi}}=e^{-i(H_L-H_R)t_\omega}\mathcal{O}_L(x)\ket{\psi_\beta}\,.
\end{equation}
Thus, the mutual information computed at $t_-=t_+=0$ in these two states has exactly the same functional dependence on $t_\omega$. This can be confirmed from our explicit formulas for $I_{A:B}$ in this section.

Notice that in the previous single sided entanglement entropy calculations, we used translation invariance to write the time dependence of the operator insertions in \eqref{eq:shift1} as a function of $t_- + t_\omega$. When computing two-sided observables, the same shift will be applied on the CFT time $t_+$ in the opposite boundary. This is consistent with the TFD path integral construction \cite{eternal} and it also appears naturally in our holographic dual model as it can explicitly be seen in the embedding equations \eqref{MapN} and \eqref{MapN1} appearing in appendix B.

\subsection{Semi-infinite intervals}\label{ssi}
Before proceeding with finite entangling regions, consider $A$ and $B$ to be semi-infinite intervals $x\in [0,\infty)$. We want to clarify the difference between previous results for the second ($n=2$) Renyi mutual information in this setup \cite{Caputa:2014eta} and our current mutual information $(n=1)$ discussion. In the large central charge $c$ limit, and after the insertion of a local operator, the second Renyi entanglement entropy of the union $S^{(2)}_{A\cup B}$ grows linearly with time. Equivalently, the change in the second Renyi mutual information for semi-infinite intervals decreases linearly with time \cite{Caputa:2014eta}
\begin{equation}
\Delta I^{(2)}_{A\cup B}\simeq -\frac{8\pi h_{\psi}}{\beta}t\,.
\end{equation}
This holds for late times in the regime where $1\ll h_\psi\ll c$. Below, we want to compare this behavior with a large $c$ computation of the mutual information ($n\to 1$) with twist operators and for heavy local operators $h_\psi \sim \mathcal{O}(c)$ (as in \cite{Roberts:2014ifa}).

To compute the entanglement entropy $S_{A \cup B}$ between two semi-infinite intervals $A$ and $B$ with starting point $L_1=y>0$ on each boundary CFT, we must calculate
\begin{equation}
\Tr\rho^n_{A \cup B} (t)
=\frac{\langle\Psi(x_1,\bar{x}_1)\sigma_n(x_2,\bar{x}_2)\tilde{\sigma}_n(x_6,\bar{x}_6)\Psi^{\dagger}(x_4,\bar{x}_4)\rangle}{\left(\langle\psi(x_1,\bar{x}_1)\psi^{\dagger}(x_4,\bar{x}_4)\rangle_{C_1}\right)^n}
\end{equation}
with the insertion points
\begin{align}
x_1&=-i\epsilon,\quad x_2=y-t_\omega-t_-,\quad x_6=y+i\frac{\beta}{2}- t_+ - t_\omega,\quad x_4=+i\epsilon \nn\\
\bar{x}_1&=+i\epsilon,\quad \bar{x}_2=y+t_\omega+t_-,\quad \bar{x}_6=y-i\frac{\beta}{2}+ t_+ + t_\omega,\quad \bar{x}_4=-i\epsilon\,.
\end{align}
Notice the edge of region B (the location $x_6$) was shifted by $i\frac{\beta}{2}$, in accordance with \eqref{eq2s} and the dependence on $t_+$ is also through $t_++t_\omega$.

We follow the same strategy as before : after mapping the cylinder to a plane by $w = e^{\frac{2\pi}{\beta} x}$, it is the cross-ratio $z, \bar{z}$ on the plane that controls the relevant 4-pt function
\begin{align}
  z= \frac{w_{12} w_{64}}{w_{16} w_{24}}
&\simeq1+\frac{2\pi i \epsilon}{\beta}\frac{\cosh\frac{\pi (t_--t_+ )}
{\beta}}{\sinh\frac{\pi (y-t_--t_\omega)}{\beta}\cosh\frac{\pi (y- t_+-t_\omega)}{\beta}}+{\cal O}(\epsilon^2)\,, \\
\bar{z}= \frac{\bar{w}_{12} \bar{w}_{64}}{\bar{w}_{16} \bar{w}_{24}}
&\simeq 1-\frac{2\pi i \epsilon}{\beta}\frac{\cosh\frac{\pi (t_--t_+)}
{\beta}}{\sinh\frac{\pi (y+t_-+t_\omega)}{\beta}\cosh\frac{\pi (y+ t_+ + t_\omega)}{\beta}}+{\cal O}(\epsilon^2)\,.
\end{align}
As before, the sign of the imaginary part of $z$ flips for $t_\omega+t_- >y$ from positive to negative, whereas that of $\bar{z}$ is negative for all $t_\omega$ and $t_\pm$. To the first order in $\epsilon$, we find $(z, \bar{z}) \to (1, 1)$ for $t_\omega+t_- <y$ and
$(z, \bar{z}) \to (e^{2 \pi i}, 1)$ for $t_\omega+t_- > y$. Using the same method as for the single sided case,
we obtain the entanglement entropy $S_{A \cup B}$ for the semi-infinite intervals to be
\begin{equation}
S_{A \cup B} =
\left\{
\begin{array}{ll}
\frac{c}{3} \log \left[\frac{\beta}{\pi\varepsilon_{UV}} \cosh\frac{\pi \Delta t}{\beta} \right]
& (t_- + t_\omega < y) \\
\frac{c}{6}\log \left[ \left( \frac{\beta}{\pi\varepsilon_{UV}} \right)^2
\frac{\beta}{\pi\epsilon} \frac{\sin \pi \alpha_{\psi}}{\alpha_{\psi}}
\cosh\frac{\pi \Delta t}{\beta}
\sinh\frac{\pi (t_\omega+t_--y)}{\beta} \cosh\frac{\pi (t_+ + t_\omega -y)}{\beta} \right]
&(y< t_- +t_\omega)
\end{array}
\right.
\end{equation}
where $\Delta t = t_- - t_+$. In particular, when $t_- = t_+ = 0$ and $t_\omega$ is very large ($y \ll t_\omega$), $S_{A \cup B}$ grows linearly with $t_\omega$
\begin{align}
S_{A \cup B} \sim
\frac{\pi c}{3 \beta} (t_\omega - y) +
\frac{c}{6}\log \left[ \left( \frac{\beta}{\pi\varepsilon_{UV}} \right)^2
\frac{\beta}{4\pi\epsilon} \frac{\sin \pi \alpha_{\psi}}{\alpha_{\psi}}
\right] .
\end{align}
We can find that the first term behaves like thermal entropy which is proportional to $1/\beta$.
Equivalently the mutual information decreases linearly with $t_w$ but now with a coefficient proportional to the central charge $c$. As explained in \cite{Caputa:2014eta}, this behavior is interpreted as the destruction of the entanglement between $CFT_L$ and $CFT_R$ and the broken "entanglement bond" reconnects between the subsystem $A \cup B$ and its complement. The unit cost of this reconnection process is proportional to  $\frac{\pi c}{3 \beta}$ (which is the entropy density $S_{density}$).

\subsection{Mutual information for finite intervals}

In this section we compute the mutual information between finite regions $A$ and $B$ in opposite boundaries in the TFD state at large central charge. The setup is depicted on Fig. \ref{twoside}.\\
The mutual information is defined as
\begin{equation}
  I_{A:B} = S_A + S_B - S_{A\cup B}
\end{equation}
where $S_{A\cup B}$ stands for the entanglement entropy of the union of the two intervals. Each of the three entropies is computed using the replica trick in terms of the correlators of the local operators $\Psi$ \eqref{PSI} and twist fields inserted at the endpoints of the entangling regions. Since we already computed $S_A$, we focus on $S_B$ and $S_{A\cup B}$.

\subsubsection{\texorpdfstring{$S_B$}{S\_B}}

The calculation of the entanglement entropy $S_B$ in the second CFT$_R$ is analogous to the one for $S_A$. It involves the same type of normalised correlation function
\begin{equation}
\Tr\rho^n_B(t)=\frac{\langle\Psi(x_1,\bar{x}_1)\sigma_n(x_5,\bar{x}_5)\tilde{\sigma}_n(x_6,\bar{x}_6)\Psi^{\dagger}(x_4,\bar{x}_4)\rangle}{\left(\langle\psi(x_1,\bar{x}_1)\psi^{\dagger}(x_4,\bar{x}_4)\rangle_{C_1}\right)^n}
\end{equation}
but with the insertion points for the twist operators conveniently shifted by $\pm i \frac{\beta}{2}$ as reviewed above
\begin{eqnarray}
x_1=-i\epsilon,\quad x_5=y+L+i\frac{\beta}{2}- t_+-t_\omega,\quad x_6=y+i\frac{\beta}{2}- t_+-t_\omega,\quad x_4=+i\epsilon\nn\\
\bar{x}_1=+i\epsilon,\quad \bar{x}_5=y+L-i\frac{\beta}{2}+ t_++t_\omega,\quad \bar{x}_6=y-i\frac{\beta}{2}+ t_++t_\omega,\quad \bar{x}_4=-i\epsilon\,.
\end{eqnarray}
As stated earlier, $t_++t_\omega$ is the shifted time being used in the right CFT. To compute the 4-pt correlator, we compose the map \eqref{eq:c-plane} with
\begin{equation}
z(w)=\frac{(w_1-w)(w_6-w_4)}{(w_1-w_6)(w-w_4)}\,.
\end{equation}
The corresponding cross-ratios equal
\begin{equation}
\begin{aligned}
z=z_5&\simeq1-\frac{2\pi i \epsilon}{\beta}\frac{\sinh\frac{\pi\,L}{\beta}}{\cosh\frac{\pi (y- t_+-t_\omega)}{\beta}\cosh\frac{\pi (y+L- t_+-t_\omega)}{\beta}}+{\cal O}(\epsilon^2)\,, \\
\bar{z}=\bar{z}_5&\simeq 1+\frac{2\pi i \epsilon}{\beta}\frac{\sinh\frac{\pi\,L}{\beta}}{\cosh\frac{\pi (y+t_++t_\omega)}{\beta}\cosh\frac{\pi (y+L+ t_++t_\omega)}{\beta}}+{\cal O}(\epsilon^2)\,.
\end{aligned}
\end{equation}
Notice that the signs of the imaginary parts are the same for all $t_+$ and $t_w$. Thus, in the small $\epsilon$ limit, $(z,\bar{z})\to (1,1)$ for all $t_+$. This reflects the intuition that the local perturbation turned on on the left CFT has no effect, at lowest order in $\epsilon$, in the quantum entanglement measured in the right CFT. Using the expansion in these cross-ratios, we derive
\begin{equation}
\boxed{
S_B=\frac{c}{3}\log\left(\frac{\beta}{\pi \varepsilon_{UV}}\sinh\frac{\pi L}{\beta}\right) \quad \quad \forall\,t_+
}\label{eq:S_B_CFT}
\end{equation}
Thus, quantum entanglement in the region $B$ remains thermal for all $t_+$ at lowest order in $\epsilon$, i.e. $\Delta S_B = 0$.

\subsubsection{\texorpdfstring{$S_{A\cup B}$}{S\_\{A∪B\}}}
The most interesting piece in the mutual information is $S_{A\cup B}$. Following \cite{Mor}, this requires the calculation of the 6-pt function
\begin{equation}
\Tr\rho^n_{A\cup B}(t)=\frac{\langle\psi(x_1,\bar{x}_1)\sigma_n(x_2,\bar{x}_2)\tilde{\sigma}_n(x_3,\bar{x}_3)\sigma_n(x_5,\bar{x}_5)\tilde{\sigma}_n(x_6,\bar{x}_6)\psi^{\dagger}(x_4,\bar{x}_4)\rangle}{\left(\langle\psi(x_1,\bar{x}_1)\psi^{\dagger}(x_4,\bar{x}_4)\rangle_{C_1}\right)^n}
\label{eq:sab}
\end{equation}
where the different insertion points correspond to the different interval endpoints
\begin{eqnarray}
x_1&=&-i\epsilon,\quad x_2=y-t_--t_\omega,\quad x_3=y+L-t_--t_\omega,\quad x_4=+i\epsilon\nn\\
\bar{x}_1&=&+i\epsilon,\quad \bar{x}_2=y+t_-+t_\omega,\quad \bar{x}_3=y+L+t_-+t_\omega,\quad \bar{x}_4=-i\epsilon\nn\\
x_5&=&y+L+i\frac{\beta}{2}- t_+-t_\omega,\quad x_6=y+i\frac{\beta}{2}- t_+-t_\omega,\nn\\
\bar{x}_5&=&y+L-i\frac{\beta}{2}+ t_++t_\omega,\quad \bar{x}_6=y-i\frac{\beta}{2}+ t_++t_\omega\,.
\end{eqnarray}
See Appendix \ref{ap-A}, for further comments on the ordering of the twist operators appearing in \eqref{eq:sab}. Following the same strategy as before, we compose the two maps
\begin{equation}
w(x)=e^{\frac{2\pi}{\beta}x} \quad \text{and} \quad z(w)=\frac{(w_1-w)w_{34}}{w_{13}(w-w_4)}\,,
\end{equation}
and use the transformation properties of primary operators, to write the trace \eqref{eq:sab} as
\begin{equation}
\Tr \rho_{A\cup B}^n=\left|\frac{\beta}{\pi\varepsilon_{UV}}\sinh\left(\frac{\pi L}{\beta}\right)\right|^{-8H_\sigma}|1-z|^{4H_\sigma}\left|z_{56}\right|^{4H_\sigma}\langle\psi|\sigma_n(z,\bar{z})\tilde{\sigma}_n(1,1)\sigma_n(z_5,\bar{z}_5)\tilde{\sigma}_n(z_6,\bar{z}_6)|\psi\rangle\label{main}
\end{equation}
where the cross-ratios $(z,\bar{z})$ are given in \eqref{eq:zA}, and $z_i\equiv z(w_i)$.

In the following, we discuss two different CFT channels: S and T-channel, where we compute this 6-pt function on the plane in the large $c$ limit (see a detailed discussion in \cite{Hartman:2013mia}). The corrections to the particular channel choice are suppressed by $e^{-O(c)}$ factors. We will explicitly see how these channels match the two different bulk geodesics determining the holographic entanglement entropy in our holographic discussions. The upshot is that S and T-channel correspond to the disconnected and connected geodesics for the holographic calculation of $S_{A\cup B}$, respectively.

\paragraph{S-channel :} Let us introduce a resolution of the identity
\begin{equation}
\langle\psi|\sigma_n(z,\bar{z})\tilde{\sigma}_n(1,1)\sigma_n(z_5,\bar{z}_5)\tilde{\sigma}_n(z_6,\bar{z}_6)|\psi\rangle=\sum_{\alpha}\langle\psi|\sigma_n(z,\bar{z})\tilde{\sigma}_n(1,1)\ketbra{\alpha}\sigma_n(z_5,\bar{z}_5)\tilde{\sigma}_n(z_6,\bar{z}_6)|\psi\rangle
\end{equation}
where the sum runs over all possible intermediate states.

Consider the first 4-pt function $\langle\psi|\sigma_n(z,\bar{z})\tilde{\sigma}_n(1,1)\ket{\alpha}$. As we have seen in \eqref{eq:zA}, the relevant limit corresponding to $\ep\to 0$, is either $(z,\bar{z})\to (1,1)$ or $(z,\bar{z})\to (e^{2\pi i},1)$. Thus, in either limit, the correlation function can be computed using the OPE of twist operators \cite{Calabrese:2010he,Perlmutter:2013paa}
\begin{equation}
  \sigma_n(z,\bar{z})\tilde{\sigma}_n(1,1) \sim \mathbb{I} + {\cal{O}}\left((z-1)^r\right) \quad \quad r\in \mathbb{Z}^+
\label{eq:OPE}
\end{equation}
Ignoring the terms proportional to $(z-1)$ and focusing in the dominant contribution due to the identity operator, we reach the important conclusion that the summation over the entire set of intermediate steps is restricted to $\ket{\alpha}=\ket{\psi}$ due to the orthogonality of 2-pt functions in any CFT. We stress that we could have reached the same conclusion in the limit of small $L_2-L_1$, but this is {\it not} required in our set-up.

Thus, our 6-pt function can then be approximated by
\begin{equation}
  \langle\psi|\sigma_n(z,\bar{z})\tilde{\sigma}_n(1,1)\sigma_n(z_5,\bar{z}_5)\tilde{\sigma}_n(z_6,\bar{z}_6)|\psi\rangle\simeq   \langle\psi|\sigma_n(z,\bar{z})\tilde{\sigma}_n(1,1)\ket{\psi}\,\bra{\psi}\sigma_n(z_5,\bar{z}_5)\tilde{\sigma}_n(z_6,\bar{z}_6)|\psi\rangle
\end{equation}
The first 4-pt function equals $G(z,\bar{z})$ in \eqref{eq:mcor}, whereas the second 4-pt function factor will be proportional to the same function but evaluated at a different cross-ratio. To see this, consider the map taking $z_1\to0$ and $z_4\to \infty$
\begin{equation}
\tilde{z}(x)=\frac{(z_1-x)(z_6-z_4)}{(z_1-z_6)(x-z_4)}\,.
\end{equation}
This allows us to write the desired correlator as
\begin{equation}
\bra{\psi}\sigma_n(z_5,\bar{z}_5)\tilde{\sigma}_n(z_6,\bar{z}_6)|\psi\rangle=\left|1-\tilde{z}_5\right|^{4H_\sigma}|z_{56}|^{-4H_\sigma}\bra{\psi}\sigma_n(\tilde{z}_5,\bar{\tilde{z}}_5)\tilde{\sigma}_n(1,1)|\psi\rangle\,.
\end{equation}
Thus, the leading contribution in this channel is
\begin{eqnarray}
\Tr \rho_{A\cup B}^n\simeq \left|\frac{\beta}{\pi\varepsilon_{UV}}\sinh\left(\frac{\pi L}{\beta}\right)\right|^{-8H_\sigma}|1-z|^{4H_\sigma}\left|1-\tilde{z}_5\right|^{4H_\sigma}G(z,\bar{z})G(\tilde{z}_5,\bar{\tilde{z}}_5)+...
\end{eqnarray}
where the dots stand for the contributions coming from the subleading terms in the OPE of the twist operators \eqref{eq:OPE}.
Interestingly, since the cross-ratio $\tilde{z}_5$ equals $z_5$, the cross-ratio determining $S_B$, we reach the conclusion that
\begin{equation}
S_{A\cup B}=S_A+S_B,\qquad \text{and}\qquad
I_{A:B}=0\,.
\end{equation}
This channel reproduces the bulk expectation coming from geodesics joining points in the same boundary, leading to a vanishing mutual information.

\paragraph{T-channel :} We could also introduce the resolution of the identity as follows
\begin{equation}
\langle\psi|\sigma_n(z,\bar{z})\tilde{\sigma}_n(1,1)\sigma_n(z_5,\bar{z}_5)\tilde{\sigma}_n(z_6,\bar{z}_6)|\psi\rangle=\sum_{\alpha}\langle\psi|\sigma_n(z,\bar{z})\tilde{\sigma}_n(z_6,\bar{z}_6)\ketbra{\alpha}\sigma_n(z_5,\bar{z}_5)\tilde{\sigma}_n(1,1)|\psi\rangle\,.
\end{equation}
Notice the correlations involve twist operators inserted in different boundaries. Thus, we expect this channel to reproduce the bulk contribution from geodesics connecting both boundaries. Remember that in the small $\epsilon$ limit, we already argued that $z_5\to 1$. Thus, we can use the same OPE argument as above to conclude that the dominant contribution comes from $\ket{\alpha}=\ket{\psi}$. By definition, this gives
\begin{equation}
\bra{\psi}\sigma_n(z_5,\bar{z}_5)\tilde{\sigma}_n(1,1)|\psi\rangle=G(z_5,\bar{z}_5)
\end{equation}
The remaining correlation is again proportional to the same function, but evaluated at a different cross-ratio. This is proved by considering the map
\begin{equation}
\tilde{z}(x)=\frac{(z_1-x)(z_6-z_4)}{(z_1-z_6)(x-z_4)}\,.
\end{equation}
which allows us to derive
\begin{equation}
\langle\psi|\sigma_n(z,\bar{z})\tilde{\sigma}_n(z_6,\bar{z}_6)\ket{\psi}=|1-\tilde{z}_2|^{4H_\sigma}|z_{26}|^{-4H_\sigma}G(\tilde{z}_2,\bar{\tilde{z}}_2)
\end{equation}
where $z_2=z(w_2)=z$ as in \eqref{eq:zA} and $\tilde{z}_2=\tilde{z}(z_2)$. Thus, after some manipulations we have
\begin{equation}
\begin{aligned}
\Tr \rho_{A\cup B}^n &
& \simeq \left|\frac{\beta}{\pi\varepsilon_{UV}}\sinh\left(\frac{\pi L}{\beta}\right)\right|^{-8H_\sigma}\left|\frac{x}{1-x}\right|^{4H_\sigma}|1-z_5|^{4H_\sigma}|1-\tilde{z}_2|^{4H_\sigma}G(\tilde{z}_2,\bar{\tilde{z}}_2)G(z_5,\bar{z}_5)+...
\end{aligned}
\end{equation}
where $(x,\bar{x})$ are the cross-ratios computed out of the insertion points of the four twist operators
\begin{equation}
x=\frac{z_{23}z_{56}}{z_{25}z_{36}}=\frac{w_{23}w_{56}}{w_{25}w_{36}}=\frac{2\sinh^2\frac{\pi\,L}{\beta}}{\cosh\frac{2\pi\,L}{\beta}+\cosh\frac{2\pi(t_- -t_+)}{\beta}}=\bar{x}\,,
\label{mainCR}
\end{equation}
what allows us to write the dominant contribution from the T-channel as
\begin{equation}
\Tr \rho^n_{A\cup B}\simeq \left|\frac{\beta}{\pi \varepsilon_{UV}}\cosh\left(\frac{\pi \Delta t}{\beta}\right)\right|^{-8H_\sigma}|1-\tilde{z}_2|^{4H_\sigma}G(\tilde{z}_2,\bar{\tilde{z}}_2)|1-z_5|^{4H_\sigma}G(z_5,\bar{z}_5)\,+...,
\end{equation}
where $\Delta t=t_--t_+$ and the cross-ratios
\begin{equation}
\begin{aligned}
z_5&=1-\frac{2\pi i \epsilon}{\beta}\frac{\cosh\frac{\pi (t_--t_+)}{\beta}}{\sinh\frac{\pi(y+L-t_--t_\omega)}{\beta}\cosh\frac{\pi(y+L-t_+-t_\omega)}{\beta}}+{\cal O}(\epsilon^2)\,, \\
\bar{z}_5&=1+\frac{2\pi i \epsilon}{\beta}\frac{\cosh\frac{\pi (t_--t_+)}{\beta}}{\sinh\frac{\pi(y+L+t_-+t_\omega)}{\beta}\cosh\frac{\pi(y+L+t_++t_\omega)}{\beta}}+{\cal O}(\epsilon^2)\,, \\
\tilde{z}_2&=1+\frac{2\pi i \epsilon}{\beta}\frac{\cosh\frac{\pi (t_--t_+)}{\beta}}{\sinh\frac{\pi(y-t_--t_\omega)}{\beta}\cosh\frac{\pi(y-t_+-t_\omega)}{\beta}}+{\cal O}(\epsilon^2)\,,\\
\tilde{\bar{z}}_2&=1-\frac{2\pi i \epsilon}{\beta}\frac{\cosh\frac{\pi (t_--t_+)}{\beta}}{\sinh\frac{\pi(y+t_-+t_\omega)}{\beta}\cosh\frac{\pi(y+t_++t_\omega)}{\beta}}+{\cal O}(\epsilon^2)\,,\label{CR52}
\end{aligned}
\end{equation}

Now, using that at large central charge and for two heavy and two light operators we have the identity \cite{Fitzpatrick:2014vua,Fitzpatrick:2015zha}
\begin{equation}
|1-z|^{4h}G(z,\bar{z})\simeq \left(\frac{z^{\frac{1-\alpha}{2}}(1-z^{\alpha})\bar{z}^{\frac{1-\alpha}{2}}(1-\bar{z}^{\alpha})}{\alpha^2(1-z)(1-\bar{z})}\right)^{-2h}
\end{equation}
as well as \eqref{CR52} we can extract the behaviour of $S_{A\cup B}$ for any time regime. Let us analyse this carefully below assuming as before that $0<y<y+L$.\\

It is clear that the monodromies of the correlator are determined depending on the relation of $t_-+t_\omega$ with $y$ and $y+L$. From \eqref{CR52} the signs of the imaginary parts of $\bar{z}_5$ and $\tilde{\bar{z}}_2$ do not change with time and we have $\bar{z}_5\simeq 1$ and $\tilde{\bar{z}}_2\simeq 1$. On the other hand $\tilde{z}_2\simeq e^{2\pi i}$ when $t_-+t_\omega>y$ and $z_5\simeq e^{-2\pi i}$ when $t_-+t_\omega>y+L$. This gives us three possible contributions:
\begin{equation}
\boxed{
\begin{aligned}
 S_{A\cup B} & \simeq \frac{2c}{3}\log \left|\frac{\beta}{\pi \varepsilon_{UV}}\cosh\left(\frac{\pi \Delta t}{\beta}\right)\right| \quad t_-+t_\omega < y \\
 S_{A\cup B} &\simeq \frac{2c}{3}\log \left|\frac{\beta}{\pi \varepsilon_{UV}}\cosh\left(\frac{\pi \Delta t}{\beta}\right)\right| \\
 &+\frac{c}{6}\log\left(\frac{\beta}{\pi\epsilon}\frac{\sin\pi \alpha_\psi}{\alpha_\psi}\frac{\sinh\frac{\pi(t_-+t_w-y)}{\beta}\cosh\frac{\pi(t_++t_w-y)}{\beta}}{\cosh\frac{\pi \Delta t}{\beta}}\right)\quad y<t_-+t_\omega<y+L
 \end{aligned}
 }
 \label{SABin}
\end{equation}
and for $t_-+t_\omega >y+L$ we can rewrite our trace as
\begin{eqnarray}
\Tr \rho^n_{A\cup B}\simeq \left|\frac{\beta}{\pi \varepsilon_{UV}}\cosh\left(\frac{\pi \Delta t}{\beta}\right)\right|^{-4H_\sigma}\left(\frac{\beta}{\pi\epsilon}\frac{\sin\pi \alpha_\psi}{\alpha_\psi}\frac{\sinh\frac{\pi(t_-+t_w-y)}{\beta}\cosh\frac{\pi(t_++t_w-y)}{\beta}}{\cosh\frac{\pi\Delta t}{\beta}}\right)^{-2H_\sigma}\nn\\
\times\left|\frac{\beta}{\pi \varepsilon_{UV}}\cosh\left(\frac{\pi \Delta t}{\beta}\right)\right|^{-4H_\sigma}\left(\frac{\beta}{\pi\epsilon}\frac{\sin\pi \alpha_\psi}{\alpha_\psi}\frac{\sinh\frac{\pi(t_-+t_w-y-L)}{\beta}\cosh\frac{\pi(t_++t_w-y-L)}{\beta}}{\cosh\frac{\pi\Delta t}{\beta}}\right)^{-2H_\sigma}\,.\label{TrAB}
\end{eqnarray}
The entanglement entropy $S_{A\cup B}$ in this time regime can then be written as
\begin{equation}
\boxed{
\begin{aligned}
 S_{A\cup B} &\simeq \frac{c}{6}\log\left(\frac{\sinh\frac{\pi(t_-+t_\omega-y)}{\beta}\cosh\frac{\pi(t_++t_\omega-y)}{\beta}}{\cosh\frac{\pi \Delta t}{\beta}}\frac{\sinh\frac{\pi(t_-+t_\omega-y-L)}{\beta}\cosh\frac{\pi(t_++t_\omega-y-L)}{\beta}}{\cosh\frac{\pi \Delta t}{\beta}}\right) \\
&+ \frac{2c}{3}\log \left|\frac{\beta}{\pi \varepsilon_{UV}}\cosh\left(\frac{\pi \Delta t}{\beta}\right)\right|+\frac{c}{3}\log\left(\frac{\beta}{\pi\epsilon}\frac{\sin\pi \alpha_\psi}{\alpha_\psi}\right) \quad \quad  t_-+t_\omega >y+L
\end{aligned}
}
\label{eq:saub}
\end{equation}
Notice that \eqref{TrAB} resembles the contributions from two different pieces. In the holographic part, these will be interpreted as the contributions from two bulk geodesics connecting points in opposite boundaries.

\subsection{The evolution of the Mutual information}
The evolution of the mutual information after turning on the local excitation can now be computed in the large central charge limit.
At early times $t_-+t_\omega<y$, the single sided entropies are thermal $S_A\simeq S_B=S_{\text{thermal}}$. Thus, the mutual information equals
\begin{equation}
I^0_{A:B}\equiv \frac{2c}{3}\log\left(\frac{\sinh\frac{\pi\,L}{\beta}}{\cosh\frac{\pi \Delta t}{\beta}}\right)\,.
\label{eq:region1}
\end{equation}
This is clearly finite and depends on the Hamiltonian driving the evolution. If we use the bulk isometry $H_L- H_R$, then $t_-=t_+$ and $\Delta t = 0$, giving rise to a time independent mutual information, as it should. Notice that positivity of the mutual information in this case requires $\pi L/\beta \gtrsim 1$. Whereas for the $H_L+H_R$ Hamiltonian, we recover the mutual information time decrease discussed in \cite{HaMa}.

These results can be understood using causality considerations : for $t_-+t_\omega<y$, the perturbation did not enter into region $A$ and could not possibly disturb the original thermal entanglement. Once the excitation reaches region $A$ $(y<t_-+t_\omega<y+L)$, using \eqref{eq:S_A_CFT}, \eqref{eq:S_B_CFT} and \eqref{SABin}, the mutual information evolves as
\begin{equation}
I_{A:B}\simeq I^0_{A:B}+ \frac{c}{6}\log\left[\frac{\sinh\frac{\pi(y+L -t_- -t_\omega)}{\beta}\cosh\frac{\pi \Delta t}{\beta}}{\cosh\frac{\pi(t_++t_w-y)}{\beta}\sinh\frac{\pi\,L}{\beta}}\right]
\label{eq:region2}
\end{equation}
Note how the dependence on the conformal dimension $h_\psi$ of the perturbation cancels, between the contributions in $S_A$ and $S_{A\cup B}$, in this regime. It would be interesting to understand the mechanism behind this large $c$ behaviour of the mutual information.

In the last region $t_- + t_\omega > y+L > y$ the mutual information equals
\begin{equation}
\boxed{
\begin{aligned}
I_{A:B} &\simeq I^0_{A:B} -\frac{c}{3}\log\left(\frac{\beta}{\pi\epsilon}\frac{\sin\pi \alpha_\psi}{\alpha_\psi}\right) \\
&-\frac{c}{6}\log\left(\frac{\sinh\frac{\pi(t_-+t_\omega-y)}{\beta}\cosh\frac{\pi(t_++t_\omega-y)}{\beta}}{\cosh\frac{\pi \Delta t}{\beta}}\frac{\sinh\frac{\pi(t_-+t_\omega-y-L)}{\beta}\cosh\frac{\pi(t_++t_\omega-y-L)}{\beta}}{\cosh\frac{\pi \Delta t}{\beta}}\right)
\end{aligned}
}\label{eq:mutual_information_CFT}
\end{equation}

It is important to stress that when extracting the answer for the mutual information for various times $t_\mp$ as well as $t_\omega$, one has to maximise the mutual information between the $S$ and the $T$ channel answers so that it is always non-negative.

\section{Scrambling time}

Shenker and Stanford \cite{Shenker:2013pqa} defined the scrambling time $t^\star_\omega$ as the time scale at which the perturbation has destroyed all the preexistent correlations. In our notation, their condition reduces to setting $t_-=t_+=0$ and to study the vanishing of the mutual information
\begin{equation}
I_{A:B}(t^\star_\omega)=0\,.
\label{eq:stime}
\end{equation}
Evaluating our previous results \eqref{eq:region1}, \eqref{eq:region2} and \eqref{eq:mutual_information_CFT} for $t_-=t_+=0$, we obtain
\begin{eqnarray}
  I_{A:B} &\simeq& \frac{2c}{3}\log \sinh\frac{\pi\,L}{\beta}\,, \quad t_\omega < y \\
  I_{A:B} &\simeq & \frac{c}{6}\log\left[\frac{\left(\sinh\frac{\pi\,L}{\beta}\right)^3\sinh \frac{\pi(y+L-t_\omega)}{\beta}}{\cosh\frac{\pi(t_\omega-y)}{\beta}}\right]\,, \quad y < t_\omega < y+L \\
  I_{A:B} &\simeq & \frac{2c}{3}\log \sinh\frac{\pi\,L}{\beta} -\frac{c}{3}\log\left(\frac{\beta}{\pi\epsilon}\frac{\sin\pi \alpha_\psi}{\alpha_\psi}\right) \nonumber \\
  & & - \frac{c}{6}\log \left(\frac{\sinh \frac{2\pi(t_\omega-y)}{\beta}\,\sinh \frac{2\pi(t_\omega - y-L)}{\beta}}{4}\right)\,, \quad t_\omega > y+L
\end{eqnarray}
Notice the mutual information is a monotonically decreasing function of $t_\omega$. Thus, starting with a positive mutual information, i.e.  $\pi L/\beta \gtrsim 1$, there is a single root $t^\star_\omega$ where \eqref{eq:stime} holds. After that, by switching channels, the mutual information remains zero.

The first question to answer is whether $t^\star_\omega \in (y,\,y+L)$ or whether $t^\star_\omega > y+L$. Clearly, the second condition can only hold if the mutual information is positive at the transition. This requirement gives rise to the constraint
\begin{equation}
  t^\star_\omega > y+L \,\, \Rightarrow \,\, I_{A:B}(y + L + \epsilon) > 0 \quad \Rightarrow \quad \frac{\beta}{\pi \epsilon} \left( \frac{\sin \pi \alpha_\psi}{\alpha_\psi} \right)^2 < \frac{\sinh^3\frac{\pi L}{\beta}}{\cosh\frac{\pi L}{\beta}}\,,
\label{eq:rscrambling}
\end{equation}
where we already used $\epsilon\ll \beta$, as in our previous CFT analysis\footnote{If $\epsilon\ll\beta$ breaks down, then condition \eqref{eq:rscrambling} is modified.}. Because of working in this region of parameter space $(\epsilon\ll \beta)$, we conclude that only small perturbations $(\alpha_\psi\to 1)$ allow scrambling time scales $t^\star_\omega > y+L$.
Since the function $\sin (\pi\alpha_\psi)/\alpha_\psi$ is monotonically decreasing in $\alpha_\psi$ (or increasing in $h_\psi$), the smaller the perturbation is, the easier it is to fulfil condition \eqref{eq:rscrambling} for generic values of $L/\beta$. Since this is the regime considered in \cite{Shenker:2013pqa}, we will study the scrambling time under these circumstances\footnote{One can equally study the conditions under which $t^\star_\omega \in (y,\,y+L)$. These generically require heavier perturbations. We do not understand this regime, which appears precisely when the mutual information \eqref{eq:region2} is $\alpha_\psi$ independent.}.

For $t_\omega > y+L$ and $\Delta t=0$, the mutual information \eqref{eq:mutual_information_CFT} becomes
\begin{equation}
I_{A:B}(t_\omega) \simeq \frac{c}{6}\log\frac{\sinh^4\frac{\pi L}{\beta}\left(\frac{\beta}{\pi\epsilon}\frac{\sin\pi \alpha_\psi}{\alpha_\psi}\right)^{-2}}{\cosh\left(\frac{\pi (t_\omega-y)}{\beta}\right)\sinh\frac{\pi (t_\omega-y)}{\beta}\sinh\frac{\pi(t_\omega-y-L)}{\beta}\cosh\frac{\pi (t_\omega-y-L)}{\beta}}
\end{equation}
This vanishes when
\begin{equation}
 \left(\frac{2\pi \alpha_\psi \epsilon \sinh^2 \left( \frac{\pi L}{\beta} \right) }{\beta \sin(\pi\alpha_\psi) }\right)^2
  = \sinh^2 \frac{2\pi (t_\omega^\star-y)}{\beta} \cosh \frac{2 \pi L}{\beta}\left(1-\frac{\tanh\frac{2\pi L}{\beta}}{\tanh\frac{2\pi (t_\omega^\star-y)}{\beta}}\right)\,.
\label{eq:exact}
\end{equation}
Notice that $t_\omega^\star - y > L$ guarantees the positivity of the left hand side. Condition \eqref{eq:exact} gives rise to a quadratic equation in $\sinh^2 \frac{2\pi (t_\omega^\star-y)}{\beta}$. It can be shown that there is a {\it unique} consistent root, in agreement with our previous arguments. In the limit $t^\star_\omega/\beta \gg 1$, this root reduces the scrambling time $t_\omega^\star$ to
\begin{equation}
\boxed{
t^\star_\omega=y + \frac{L}{2}-\frac{\beta}{2\pi}\log\left(\frac{\beta}{\pi\epsilon}\frac{\sin\pi \alpha_\psi}{\alpha_\psi}\right)+\frac{\beta}{\pi}\log\left(2 \sinh\frac{\pi L}{\beta}\right)\label{SCRtime}
}
\end{equation}
Due to the non-compactness of the 2d CFT, no recurrences were expected to be seen in our calculation. Working in the small $h_\psi/c$ limit, as required by our analysis, then
\begin{equation}
\frac{\beta}{\pi \epsilon}\frac{\sin \pi \alpha_\psi}{\alpha_\psi}\sim \frac{E_\psi}{S_{\text{density}}},\qquad \text{where} \qquad
S_{\text{density}} =\frac{\pi c}{3\beta},
\end{equation}
and $E_\psi=\f{\pi h_{\psi}}{\ep}$ is the total energy of our local excitation given by integrating the energy density as in \cite{Caputa:2014vaa}. In this limit, the scrambling time reduces to
\begin{equation}
t^\star_\omega=y + \frac{L}{2} +\frac{\beta}{2\pi}\log \left(\frac{\pi S_{density}}{4 E_\psi}\right)
+\frac{\beta}{\pi}\log\left(2 \sinh\frac{\pi L}{\beta}\right)\,.
\end{equation}
The $\log S$ dependence is indeed consistent with the original scrambling conjecture \cite{Sekino:2008he,Shenker:2013pqa}.

\section{Holographic description}

In this section we compute the mutual information using the AdS$_3$ gravity dual of the 2d CFT set-up discussed in the previous section.
The starting thermal state $\rho_\beta$ in the quantum theory is described in the semiclassical approximation of the correspondence by a black hole in the gravity side, the BTZ black hole \cite{BTZ}. Observables involving a single Hilbert space are described by the metric outside of the event horizon, whereas observables depending on both Hilbert spaces require the Kruskal-like extension of the BTZ black hole, as described in \cite{eternal}.

To compute the time evolution of the holographic entanglement entropy, we approximate the local CFT perturbation with conformal dimension $\Delta(=2h_\psi)$
by a bulk free falling massive point particle with mass $m=\Delta/R$ \cite{NNT}. Next, we compute its back-reaction on the BTZ background in Kruskal coordinates using the coordinate transformation in \cite{Horowitz:1999gf}. Finally, we compute the entanglement entropy and the mutual information using the holographic prescription \cite{RTorig,Hubeny:2007xt}.

In three dimensions, the back-reacted metric of a point particle at $r=0$ in global coordinates is known
\begin{equation}
ds^2=-\left(r^2+R^2-\mu\right)d\tau^{2}+\frac{R^2dr^2}{r^2+R^2-\mu}+r^2d\varphi^{2}\,,
\label{metBR}
\end{equation}
where the mass of the point particle is related to $\mu$ by $\mu=8G_NR^2m=\frac{24h_\psi}{c}R^2$ and $R$ is the radius of AdS$_3$. Depending on the mass of the particle, the background describes a conical singularity or a BTZ black hole.

Now, the holographic entanglement entropy is given by $\f{R}{4G_N}=\f{c}{6}$ times the length $L_{\gamma}$ of geodesic $\gamma$ which connects the boundaries of the subsystem $A$ for which we define the entanglement entropy $S_A$. In the above metric, the entanglement entropy of the boundary region A with endpoints $(r^{(1)}_\infty,\tau^{(1)}_{\infty},\varphi^{(1)}_{\infty})$ and $(r^{(2)}_\infty,\tau^{(2)}_{\infty},\varphi^{(2)}_{\infty})$ is \cite{NNT}
\begin{equation}
S_A=\frac{c}{6}\log\left[\frac{2r^{(1)}_\infty\cdot r^{(2)}_{\infty}}{R^2}\frac{\cos\left(|\Delta \tau_{\infty}|a\right)-\cos\left(|\Delta \varphi_{\infty}|a\right)}{a^2}\right],\label{EEbr}
\end{equation}
where $a\equiv \sqrt{1-\frac{\mu}{R^2}}=\alpha_\psi$ carries the information on the perturbation, as in the CFT discussion,  $\Delta \tau_{\infty}=\tau^{(2)}_{\infty}-\tau^{(1)}_{\infty}$ and $\Delta \varphi_{\infty}=\varphi^{(2)}_{\infty}-\varphi^{(1)}_{\infty}$ satisfies $0<|\Delta \varphi_\infty|<\pi$.

Mapping the static $r=0$ geodesic to one starting at some distance $\epsilon$ from the boundary and falling into the horizon afterwards can approximate the local perturbation turned on in the boundary theory. This is precisely the approach followed in \cite{Caputa:2014eta} to describe the time dependent evolution of entanglement entropy in the bulk for locally perturbed thermal states. To describe the evolution across the horizon, which is required to study two sided correlation functions, one must use Kruskal coordinates. This is one of the tasks we will undertake in this section.

\subsection{Free falling particle in Kruskal coordinates}

The geodesic of a free falling particle in the AdS-Schwarzschild patch of the BTZ black hole
\begin{equation}
ds^2=\frac{R^2}{z^2}\left[-\left(1-Mz^2\right)dt^2_{-}+\frac{dz^2}{1-Mz^2}+d\theta^2\right]\,, \quad \quad \theta\sim \theta + 2\pi
\label{AdSS}
\end{equation}
was already computed in \cite{Caputa:2014eta} to be
\begin{equation}
t_-=\tilde{\tau},\quad \theta=0,\quad 1-Mz^2=(1-M\epsilon^2)\cosh^{-2}\left(\sqrt{M}(\tilde{\tau}+t_\omega)\right)\,.
\label{Soltw}
\end{equation}
The only addition in the expression above is the shift $\tilde{\tau}\to \tilde{\tau} + t_\omega$ to account for the initial boundary condition
$z(-t_\omega)=\epsilon$ guaranteeing the particle's energy
\begin{equation}
  E = \frac{m\,R}{\epsilon}\,\sqrt{1-M\epsilon^2}\,,
\end{equation}
matches the energy of the CFT perturbation in the small $\epsilon$ limit\footnote{In the absence of matter fields and working in the large $c$ limit, as we do, the satisfaction of this condition guarantees that our holographic model should capture the bulk description of the identity conformal block, which in 2d includes the stress tensor.}. We extend this result to the entire eternal black hole by working in Kruskal coordinates.

One way to achieve this goal is to map the global AdS$_3$ description \eqref{metBR} to Kruskal coordinates. A second one is to solve the geodesic equation directly in Kruskal coordinates. We check below that, as expected, both approaches agree.

\paragraph{Free falling particle in Kruskal coordinates :}

The Kruskal extension of the BTZ metric \eqref{AdSS} is given by
\begin{equation}
ds^2=R^2\frac{-4dudv+(-1+uv)^2d\phi^2}{(1+uv)^2}=R^2\frac{-4dT^2+4dX^2+\left(1-T^2+X^2\right)^2d\phi^2}{\left(1+T^2-X^2\right)^2}\,,
\label{eq:kruskalm}
\end{equation}
where $u=T-X\in \mathbb{R}$, $v=T+X\in\mathbb{R}$ with their range satisfying $-1< uv<1$ and $\phi\sim \phi+4\pi^2/\beta$. The conformal boundary, horizons and singularities are at $uv=-1$, $uv=0$ and $uv=1$, respectively, with the left and right Kruskal regions defined by
\begin{equation}
\begin{aligned}
\text{Left:}\quad R_-&=\{0\le u,-1\le uv\le 0\}\\
\text{Right:}\quad R_+&=\{u\le 0,-1\le uv\le 0\}
\end{aligned}
\end{equation}
Both coordinate systems  are related to the AdS-Schwarzschild patches via
\begin{equation}
\begin{aligned}
u&=\pm\sqrt{\frac{z_H-z}{z_H+z}}e^{t_{\mp}/z_H}\qquad\qquad\qquad\quad v=\mp\sqrt{\frac{z_H-z}{z_H+z}}e^{-t_{\mp}/z_H}\\
T&=\pm\sqrt{\frac{1-\sqrt{M}z}{1+\sqrt{M}z}}\sinh\left(\sqrt{M}t_\mp\right)\qquad X=\mp\sqrt{\frac{1-\sqrt{M}z}{1+\sqrt{M}z}}\cosh\left(\sqrt{M}t_\mp\right)\label{TX}
\end{aligned}
\end{equation}
Using these maps \eqref{TX}, we can rewrite the geodesic \eqref{Soltw} in the parametric form
\begin{equation}
\begin{aligned}
X(\tilde{\tau})&=-\frac{\sqrt{1-M\epsilon^2}\frac{\cosh\left(\sqrt{M}\tilde{\tau}\right)}{\cosh\left(\sqrt{M}(\tilde{\tau}+t_\omega)\right)}}{1+\sqrt{1-(1-M\epsilon^2)\cosh^{-2}\left(\sqrt{M}(\tilde{\tau}+t_\omega)\right)}}\,,\\
T(\tilde{\tau})&=\frac{\sqrt{1-M\epsilon^2}\frac{\sinh\left(\sqrt{M}\tilde{\tau}\right)}{\cosh\left(\sqrt{M}(\tilde{\tau}+t_\omega)\right)}}{1+\sqrt{1-(1-M\epsilon^2)\cosh^{-2}\left(\sqrt{M}(\tilde{\tau}+t_\omega)\right)}}\,,
\label{TXL}
\end{aligned}
\end{equation}
The initial condition $(t_-,z)=(-t_\omega,\epsilon)$ is mapped to
\begin{eqnarray}
(T_0,X_0)&=&\sqrt{\frac{1-\sqrt{M}\epsilon}{1+\sqrt{M}\epsilon}}\left(-\sinh\left(\sqrt{M}t_\omega\right),-\cosh\left(\sqrt{M}t_\omega\right)\right)\,.
\end{eqnarray}
This allows us to determine $t_\omega=t_\omega(T_0,X_0)$. Similarly, $\tilde{\tau}=\tilde{\tau}(T,X)$ can be determined from  \eqref{TX}. Altogether, we can solve for $T(X)$ as
\begin{eqnarray}
T(X)&=&-\frac{\sinh\left(\sqrt{M}t_\omega\right)}{\sqrt{1-M\epsilon^2}}\pm\sqrt{\left(X+\frac{\cosh\left(\sqrt{M}t_\omega\right)}{\sqrt{1-M\epsilon^2}}\right)^2-\frac{M\epsilon^2}{1-M\epsilon^2}}
\label{TF}
\end{eqnarray}
Proceeding in a similar way, we can obtain the geodesic $v=v(u)$ that reduces to
\begin{equation}
v(u)=-\frac{a_1 u-1}{u+a_2},\label{Solv}
\end{equation}
with
\begin{equation}
a_1=\frac{1-u_0v_0}{2u_0}=\frac{e^{\sqrt{M}t_\omega}}{\sqrt{1-M\epsilon^2}},\quad a_2=\frac{1-u_0v_0}{2v_0}=-\frac{e^{-\sqrt{M}t_\omega}}{\sqrt{1-M\epsilon^2}}\,.
\end{equation}

\begin{figure}[!ht]
  \centering
  \includegraphics[natwidth=10cm, width=10cm]{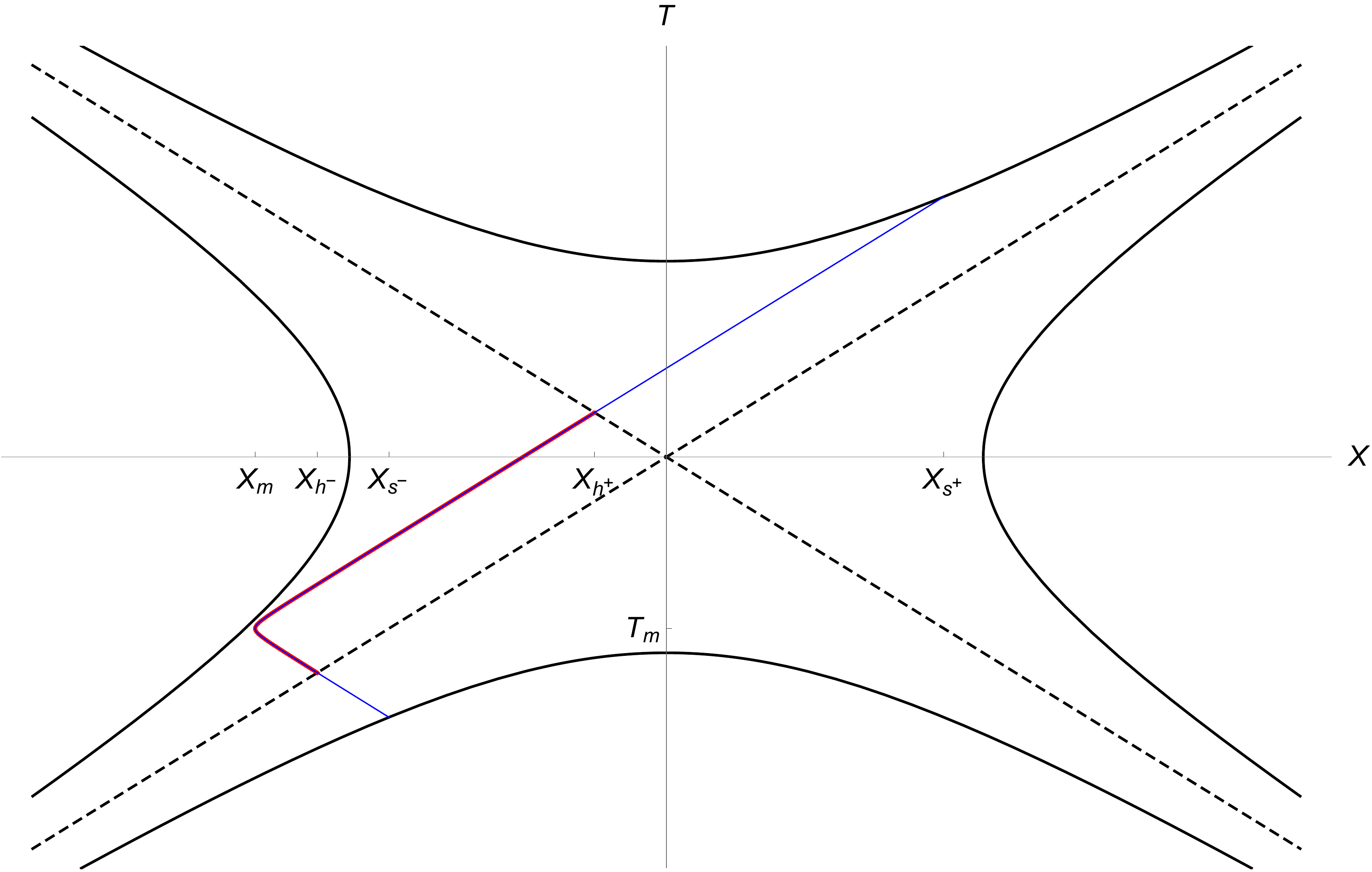}
  \caption{Plot shows our time-like geodesic on Kruskal diagram. The red part is given by \eqref{TXL} and the full geodesic (blue) by \eqref{TF}. Plot for $M=10$, $\epsilon=0.01$ and $t_\omega=0.25$}\label{KrTX}
\end{figure}

\paragraph{Checking equations of motion :} Consider the relativistic action for a particle of mass $m$ moving in the background metric \eqref{eq:kruskalm} at constant $\phi$. Working in the gauge where the parameter along the curve equals $u$, this effective action reduces to
\begin{equation}
S=-2mR\int\frac{\sqrt{v'}du}{1+uv}\,.
\end{equation}
Its equation of motion
\begin{equation}
v''(uv+1)-2v'(uv'-v)=0\,,
\end{equation}
has general solution
\begin{equation}
v(u)=\frac{C_2+\left(C_1+C^2_2\right) u}{1+C_2 u},\qquad v'(u)=\frac{C_1}{(1+C_2 u)^2}\,.
\label{SolV}
\end{equation}
This agrees with the geodesic solution \eqref{Solv} if the integration constants are matched as
\begin{equation}
C_1=-\frac{v_0(u_0v_0+1)^2}{u_0(u_0v_0-1)^2}=M\epsilon^2e^{2\sqrt{M}t_\omega},\qquad C_2=\frac{2v_0}{1-u_0v_0}=-\sqrt{1-M\epsilon^2}e^{\sqrt{M}t_\omega}\,.
\end{equation}
In $T,X$ coordinates these extended geodesics have two branches which meet at the point
\begin{equation}
X_m=\frac{\sqrt{M}\epsilon-\cosh\left(\sqrt{M}t_\omega\right)}{\sqrt{1-M\epsilon^2}},\qquad T_m=-\frac{\sinh\left(\sqrt{M}t_\omega\right)}{\sqrt{1-M\epsilon^2}}\,.
\end{equation}
They cross the future and past horizons at
\begin{equation}
X_{h^{\pm}}=-\frac{1}{2}\sqrt{1-M\epsilon^2}e^{\mp\sqrt{M}t_\omega}
\end{equation}
and hit the past and future singularities at
\begin{equation}
X_{s^{\pm}}=\pm\sinh\left(\sqrt{M}t_\omega\right)
\end{equation}
All these features can be seen in the sample geodesic plotted on the $T-X$ Kruskal diagram (Fig.\ref{KrTX}).

\subsection{Back-reacted metric}
The back reaction of the free falling particle in Kruskal coordinates can be obtained by following \cite{Horowitz:1999gf} and rewriting the metric \eqref{metBR} in Kruskal coordinates, but taking into the account the specific initial conditions discussed above. To solve this problem when $t_\omega=0$, in \cite{Caputa:2014eta}, we considered a boost in the plane $X_1-X_3$ with rapidity $\lambda_2=\lambda_2(M,\epsilon)$ . In order to introduce the further parameter $t_\omega$, we will consider a preliminary boost in $X_0-X_3$, since this corresponds to the natural boost action on the light-like coordinates $u-v$ which captures the blue-shift of energy near the horizon stressed in  \cite{Shenker:2013pqa}. In practice we then apply two particular boosts into the embedding coordinates of $AdS_3$ such that the identification between global and Kruskal coordinates becomes
\begin{equation}
\begin{aligned}
\sqrt{R^2+r^2}\sin\tau&=\cosh \lambda_1\, X_0+\sinh \lambda_1\, X_3 \\
&=R\frac{e^{\lambda_1}u+e^{-\lambda_1}v}{1+uv} \\
\sqrt{R^2+r^2}\cos\tau&=\cosh\lambda_2\,X_1-\sinh\lambda_2\,\left(\sinh \lambda_1\,X_0+\cosh \lambda_1\,X_3\right)\\
&=\frac{R\cosh\lambda_2(1-uv)}{1+uv}\left(\cosh\phi-\tanh\lambda_2\frac{e^{\lambda_1}u-e^{-\lambda_1}v}{1-uv}\right) \\
r\sin\varphi&=X_2=R\frac{1-uv}{1+uv}\sinh\phi \\
r\cos\varphi&=-\sinh\lambda_2\,X_1+\cosh\lambda_2\left(\sinh \lambda_1\,X_0+\cosh \lambda_1\,X_3\right)\\
&=\frac{R\cosh\lambda_2(1-uv)}{1+uv}\left(\frac{e^{\lambda_1}u-e^{-\lambda_1}v}{1-uv}-\tanh\lambda_2\cosh\phi\right)
\end{aligned}
\end{equation}
Solving for $r=r(u,v,\phi)$
\begin{equation}
r=\left|\frac{R(1-uv)\cosh\lambda_2}{1+uv}\right|\sqrt{\frac{\sinh^2\phi}{\cosh^2\lambda_2}+\left(\frac{e^{\lambda_1}u-e^{-\lambda_1}v}{1-uv}-\tanh\lambda_2\cosh\phi\right)^2}
\end{equation}
we can determine $\lambda_1$ and $\lambda_2$ requiring that the location of the static particle in global AdS$_3$, $r=0$, gets mapped into the free falling geodesic $v(u)$ in \eqref{Solv}. This fixes both boost parameters to be
\begin{equation}
  \lambda_1=\sqrt{M}t_\omega\,, \quad \quad  \tanh\lambda_2=\sqrt{1-M\epsilon^2}\,.
\end{equation}

These boost parameters determine the explicit map between global AdS$_3$ and a free falling particle in Kruskal coordinates :
\begin{eqnarray}
r=\frac{R}{\sqrt{M}\epsilon}\left|\frac{1-uv}{1+uv}\right|\sqrt{M\epsilon^2\sinh^2\phi+\left(\frac{e^{\sqrt{M}t_\omega}\,u-e^{-\sqrt{M}t_\omega}\,v}{1-uv}-\sqrt{1-M\epsilon^2}\cosh\phi\right)^2}\,
\end{eqnarray}
and also
\begin{eqnarray}
 \tan\tau&=&\sqrt{M}\epsilon\frac{\frac{e^{\sqrt{M}t_\omega}u+e^{-\sqrt{M}t_\omega}v}{1-uv}}{\cosh\phi-\sqrt{1-M\epsilon^2}\,\frac{e^{\sqrt{M}t_\omega}\,u-e^{-\sqrt{M}t_\omega}\,v}{1-uv}}\,, \\
\tan\varphi&=&\sqrt{M}\epsilon\frac{\sinh\phi}{\frac{e^{\sqrt{M}t_\omega}u-e^{-\sqrt{M}t_\omega}v}{1-uv}-\sqrt{1-M\epsilon^2}\cosh\phi}\,.
\label{MAPStw}
\end{eqnarray}
Using this map, we can compute the exact back-reacted metric corresponding to a free falling particle in the eternal black hole satisfying the initial condition $(u_0,\,v_0)$.

Our analysis is valid for any value of $t_\omega$. This allows us to compare with some approaches in the literature where the back reaction of the local perturbation in the CFT was approximated by a shock-wave, i.e. a BTZ spacetime in the presence of some non-trivial stress tensor localised at the horizon. Since our approach in 3d was based on computing the explicit backreaction of some point particle moving in some geodesic into the BTZ geometry, we can study the limit $t_\omega/\beta \gg 1$ in our geodesic analysis in subsection 5.1. In particular, figure \ref{KrTX} illustrates how our particle geodesic approaches a null geodesic on the horizon for such large $t_\omega$. Thus, our back-reacted metric in this particular limit should indeed correspond to a shock-wave propagating in the BTZ background as originally described in \cite{Shenker:2013pqa}. The advantage of the shock-wave description is that it also applies in higher dimensions, whereas our finite $t_\omega$ results show the agreement between CFT and bulk computations also hold beyond this regime.

\section{Bulk mutual information}

The mutual information $I_{A:B}$ between regions $A$ and $B$ in the left and right boundaries, respectively
\begin{equation}
  I_{A:B} = S_A + S_B - S_{A\cup B}\,,
\end{equation}
can now be computed by applying \eqref{EEbr} to the three different bulk geodesics providing the relevant minimal surface computing entanglement entropy in the bulk. All we need to know are the locations of the endpoints in the limits of small $\epsilon$ and $z_\infty$ that we will insert into \eqref{EEbr}. \\
To keep the gravity formulas compact, the endpoints of the intervals will be denoted by $L_i$, $i\in\{1,2\}$, where $0<L_1<L_2$. To compare with the CFT formulas one can substitute $L_1=y$ and $L_2=y+L$.

\subsection{Geodesic in the left boundary } \label{SAGr}

The two endpoints of the entanglement region $A$ in the left boundary are $(t_-,z_1,\theta_1)=(t_-,z_\infty,L_1)$ and $(t_-,z_2,\theta_1)=(t_-,z_\infty,L_2)$. It is convenient to compute their image in global AdS$_3$ using the asymptotic maps to the right and left regions (see Appendix \ref{ap-B}). Proceeding this way, their radial coordinates in global AdS$_3$ satisfy
\begin{equation}
r^{(1)}r^{(2)}\simeq \left(\frac{R}{M\epsilon z_{\infty}}\right)^2 D_1 D_2
\end{equation}
where
\begin{equation}
D_i =|\cosh\sqrt{M}L_i-\cosh\sqrt{M}(t_-+t_\omega)|\quad i=1,2
\end{equation}
whereas the other coordinates are
\begin{eqnarray}
\tan\tau^{(i)}&\simeq&\sqrt{M}\epsilon\frac{\sinh\left(\sqrt{M}(t_-+t_\omega)\right)}{\cosh\left(\sqrt{M}L_i\right)-\cosh\left(\sqrt{M}(t_-+t_\omega)\right)}\\
\tan\varphi^{(i)}&\simeq&\sqrt{M}\epsilon\frac{\sinh\left(\sqrt{M}L_i\right)}{\cosh\left(\sqrt{M}(t_-+t_\omega)\right)-\cosh\left(\sqrt{M}L_i\right)}
\end{eqnarray}
with $i=1,2$.

The length of the geodesics depends on the value of the time argument $t_-+t_w$. When $t_-+t_w < L_1 < L_2$, then the boundary points equal
\begin{equation}
\tau^{(i)}\simeq\sqrt{M}\epsilon\frac{\sinh\left(\sqrt{M}(t_-+t_\omega)\right)}{D_i},\qquad
\varphi^{(i)}\simeq\pi-\sqrt{M}\epsilon\frac{\sinh\left(\sqrt{M}L_i\right)}{D_i}\,.
\label{eq:A1}
\end{equation}
These determine the coordinate intervals to be
\begin{equation}
\begin{aligned}
  |\Delta\tau| &\simeq \frac{\sqrt{M}\epsilon}{D_1D_2}|D_2-D_1|\sinh\sqrt{M}(t_-+t_\omega)\,, \\
  |\Delta\varphi &|\simeq\frac{\sqrt{M}\epsilon}{D_1D_2}\left|D_1\sinh\sqrt{M}L_2-D_2\sinh\sqrt{M}L_1\right|\,.
\end{aligned}
\label{eq:A2}
\end{equation}
Due to the identity
\begin{equation}
D_1D_2(|\Delta \varphi|^2-|\Delta \tau|^2)=4M\epsilon^2\sinh^2\frac{\pi \Delta L}{\beta}\,,
\end{equation}
the geodesic length is
\begin{equation}
\begin{aligned}
L_\gamma&\simeq \log\left[\frac{2r^{(1)}r^{(2)}}{R^2}\frac{\cos\left(a|\Delta \tau|\right)-\cos\left(a|\Delta \varphi|\right)}{a^2}\right]\simeq\log\left[\frac{r^{(1)}r^{(2)}}{R^2}(|\Delta \varphi|^2-|\Delta \tau|^2)\right]\\
&\simeq 2\log\left(\frac{\beta}{\pi z_\infty}\sinh\frac{\pi L}{\beta}\right)\,.
\end{aligned}
\label{eq:LA}
\end{equation}
This reproduces the thermal entanglement entropy \eqref{eq:thermal}  computed in the CFT in the same time interval once both UV cut-offs are identified $\varepsilon_{UV}=z_\infty$.\\
Similarly, when $t_-+t_w > L_2 > L_1$, the boundary points equal
\begin{equation}
\tau^{(i)}\simeq\pi-\sqrt{M}\epsilon\frac{\sinh\left(\sqrt{M}(t_-+t_\omega)\right)}{D_i},\qquad
\varphi^i\simeq\sqrt{M}\epsilon\frac{\sinh\left(\sqrt{M}L_i\right)}{D_i}\,.
\end{equation}
These are different from \eqref{eq:A1}, but give rise to the same intervals \eqref{eq:A2}. Thus, the length of the bulk geodesic joining them equals \eqref{eq:LA}. This matches our CFT again.\\
Finally, when $L_2 > t_- + t_\omega > L_1$, the boundary points are
\begin{equation}
\begin{aligned}
  \tau^{(1)}&\simeq \pi-\sqrt{M}\epsilon\frac{\sinh\sqrt{M}(t_-+t_\omega)}{D_1}\,, \quad \varphi^{(1)}\simeq \sqrt{M}\epsilon\frac{\sinh\sqrt{M}L_1}{D_1}\,, \\
  \tau^{(2)}&\simeq \sqrt{M}\epsilon\frac{\sinh\sqrt{M}(t_-+t_\omega)}{D_2}\,, \quad \varphi^{(2)}\simeq \pi-\sqrt{M}\epsilon\frac{\sinh\sqrt{M}L_2}{D_2}\,.
\end{aligned}
\end{equation}
From them we can easily get the absolute values of the intervals
\begin{equation}
\begin{aligned}
  |\Delta\tau|&\simeq \pi-\frac{\sqrt{M}\epsilon}{D_1D_2}(D_1+D_2)\sinh\sqrt{M}(t_-+t_\omega)\,, \\
  |\Delta\varphi| &\simeq\pi-\frac{\sqrt{M}\epsilon}{D_1D_2}\left(D_1\sinh\sqrt{M}L_2+D_2\sinh\sqrt{M}L_1\right)\,.
\end{aligned}
\end{equation}
Notice that in the small $\epsilon$ limit we are working on, $|\Delta\tau|$ and $|\Delta\varphi|$ are close to each other
\begin{equation}
\delta=|\Delta\varphi|-|\Delta\tau|=\frac{\sqrt{M}\epsilon}{D_1D_2}\left[(D_1+D_2)\sinh\sqrt{M}(t_-+t_\omega)-D_1\sinh\sqrt{M}L_2-D_2\sinh\sqrt{M}L_1\right]\,.
\end{equation}
This allows us to write the length of the bulk geodesic between these two boundary points as
\begin{equation}
\begin{aligned}
L_\gamma &\simeq \log\left[\frac{2r^{(1)}r^{(2)}}{R^2}\frac{\cos\left(a|\Delta \tau|\right)-\cos\left(a|\Delta \varphi|\right)}{a^2}\right]\simeq\log\left[\frac{2r^{(1)}r^{(2)}}{R^2}\frac{\sin\pi a}{a}\delta\right] \\
& \simeq\log\left[\left(\frac{\beta}{\pi z_\infty}\sinh\frac{\pi\Delta L}{\beta}\right)^2\frac{\beta}{\pi\epsilon}\frac{\sin\pi a}{a}\frac{\sinh\frac{\pi(t_\omega+t_--L_1)}{\beta}\sinh\frac{\pi(L_2-t_\omega-t_-)}{\beta}}{\sinh\frac{\pi\Delta L}{\beta}}\,.
\right]
\end{aligned}
\end{equation}
where $\Delta L=L_2-L_1$. This also perfectly matches our CFT result (\ref{eq:S_A_CFT}) after employing the Ryu-Takayanagi formula.

\subsection{Geodesic in the right boundary}

The two endpoints of the entanglement region $B$ in the right boundary are $(t_+,z_1,\theta_1)=(t_+,z_\infty,L_1)$ and $(t_+,z_2,\theta_1)=(t_+,z_\infty,L_2)$. Their radial coordinates satisfy
\begin{equation}
r^{(1)}r^{(2)}\simeq \left(\frac{R}{M\epsilon z_{\infty}}\right)^2 D_1 D_2
\end{equation}
where
\begin{equation}
D_i=|\cosh\sqrt{M}L_i+\cosh\sqrt{M}(t_++t_\omega)| \quad i=1,2
\end{equation}
whereas the other coordinates are
\begin{equation}
\tan\tau^{(i)}\simeq -\sqrt{M}\epsilon\frac{\sinh\left(\sqrt{M}(t_++t_\omega)\right)}{D_i}\,, \quad \tan\varphi^{(i)}\simeq -\sqrt{M}\epsilon\frac{\sinh\left(\sqrt{M}L_i\right)}{D_i}
\end{equation}
In this case, no matter what the value of $t_+$ is, the boundary points are identified as
\begin{equation}
\tau^{(i)}\simeq-\sqrt{M}\epsilon\frac{\sinh\left(\sqrt{M}(t_++t_\omega)\right)}{D_i}\,, \quad
\varphi^{(i)} \simeq \pi-\sqrt{M}\epsilon\frac{\sinh\left(\sqrt{M}L_i\right)}{D_i}\,.
\end{equation}
These give rise to the intervals
\begin{equation}
\begin{aligned}
  |\Delta\tau| &\simeq \frac{\sqrt{M}\epsilon}{D_1D_2}|D_1-D_2|\sinh\sqrt{M}(t_++t_\omega)\,, \\
  |\Delta\varphi| &\simeq\frac{\sqrt{M}\epsilon}{D_1D_2}\left|D_1\sinh\sqrt{M}L_2-D_2\sinh\sqrt{M}L_1\right|\,.
\end{aligned}
\end{equation}
Using the identity
\begin{equation}
D_1D_2(|\Delta \varphi|^2-|\Delta \tau|^2)=4M\epsilon^2\sinh^2\frac{\pi \Delta L}{\beta}\,,
\end{equation}
the geodesic length equals
\begin{equation}
\begin{aligned}
L_\gamma&\simeq \log\left[\frac{2r^{(1)}r^{(2)}}{R^2}\frac{\cos\left(a|\Delta \tau|\right)-\cos\left(a|\Delta \varphi|\right)}{a^2}\right]\simeq\log\left[\frac{r^{(1)}r^{(2)}}{R^2}(|\Delta \varphi|^2-|\Delta \tau|^2)\right] \\
&\simeq 2\log\left(\frac{\beta}{\pi z_\infty}\sinh\frac{\pi\Delta L}{\beta}\right)\,.
\end{aligned}
\end{equation}
This reproduces the well-known thermal answer obtained in the CFT \cite{CC}
\begin{equation}
S_B\simeq \frac{c}{3}\log\left(\frac{\beta}{\pi z_\infty}\sinh\frac{\pi\Delta L}{\beta}\right)=S_{\text{thermal}},
\end{equation}
which also agrees with the CFT expression for $S_B$ in \eqref{eq:S_B_CFT}.

\subsection{Geodesics across the horizon and Mutual Information}

 We want to compute the geodesic length between two opposite boundary points located at the same space like location but with different time labels $t_\mp$. We will describe the calculation once and apply it to the two cases of interest afterwards. The product of the radial coordinates equals
\begin{equation}
r^{(1)}r^{(2)}\simeq \left(\frac{R}{M\epsilon z_{\infty}}\right)^2 D_1 D_2
\end{equation}
where
\begin{eqnarray}
D_1&=&|\cosh\sqrt{M}L_i-\cosh\sqrt{M}(t_-+t_\omega)|\\
D_2&=&|\cosh\sqrt{M}L_i+\cosh\sqrt{M}(t_++t_\omega)|
\end{eqnarray}
where $L_i$ labels the space like location in both boundaries, i.e. either $L_1$ or $L_2$. The other coordinates for the left boundary point are
\begin{eqnarray}
\tan\tau^{(1)}&\simeq&\sqrt{M}\epsilon\frac{\sinh\sqrt{M}(t_-+t_\omega)}{\cosh\sqrt{M}L_i-\cosh\sqrt{M}(t_-+t_\omega)}\\
\tan\varphi^{(1)}&\simeq&\sqrt{M}\epsilon\frac{\sinh\sqrt{M}L_i}{\cosh\sqrt{M}(t_-+t_\omega)-\cosh\sqrt{M}L_i}\,.
\end{eqnarray}
At early times, $L_i> t_\omega$, these are given by
\begin{eqnarray}
&&\tau^{(1)}\simeq \sqrt{M}\epsilon\frac{\sinh\sqrt{M}(t_-+t_\omega)}{\cosh\sqrt{M}L_i-\cosh\sqrt{M}(t_-+t_\omega)}\\
&&\varphi^{(1)}\simeq\pi-\sqrt{M}\epsilon\frac{\sinh\sqrt{M}L_i}{\cosh\sqrt{M}L_i-\cosh\sqrt{M}(t_-+t_\omega)}
\end{eqnarray}
whereas at late times,
\begin{eqnarray}
&&\tau^{(1)}\simeq \pi-\sqrt{M}\epsilon\frac{\sinh\sqrt{M}(t_-+t_\omega)}{\cosh\sqrt{M}(t_-+t_\omega)-\cosh\sqrt{M}L_1}\\
&&\varphi^{(1)}\simeq\sqrt{M}\epsilon\frac{\sinh\sqrt{M}L_1}{\cosh\sqrt{M}(t_-+t_\omega)-\cosh\sqrt{M}L_1}
\end{eqnarray}
The remaining coordinates for the right boundary point are
\begin{eqnarray}
\tan\tau^{(2)}&\simeq&-\sqrt{M}\epsilon\frac{\sinh\sqrt{M}(t_++t_\omega)}{\cosh\sqrt{M}L_i+\cosh\sqrt{M}(t_++t_\omega)}\\
\tan\varphi^{(2)}&\simeq&-\sqrt{M}\epsilon\frac{\sinh\sqrt{M}L_i}{\cosh\sqrt{M}(t_++t_\omega)+\cosh\sqrt{M}L_i}
\end{eqnarray}
In this case, they are always given by
\begin{eqnarray}
&&\tau^{(2)}\simeq -\sqrt{M}\epsilon\frac{\sinh\sqrt{M}(t_++t_\omega)}{\cosh\sqrt{M}L_i+\cosh\sqrt{M}(t_++t_\omega)}=-\sqrt{M}\epsilon\frac{\sinh\sqrt{M}(t_++t_\omega)}{D_2}\\
&&\varphi^{(2)}\simeq \pi-\sqrt{M}\epsilon\frac{\sinh\sqrt{M}L_i}{D_2}
\end{eqnarray}

Let us compute the length of the geodesic in the early time regime $L_i> t_\omega$. In this case, the interval differences are
\begin{equation}
\begin{aligned}
 |\Delta \tau| &= |\tau^{(1)}-\tau^{(2)}| \simeq\frac{\sqrt{M}\epsilon}{D_1D_2}\left|D_2\sinh\sqrt{M}(t_-+ t_\omega)+D_1\sinh\sqrt{M}\,(t_++t_\omega)\right|\,, \\
 |\Delta \varphi| &= |\varphi^{(1)} - \varphi^{(2)}| \simeq \frac{\sqrt{M}\epsilon}{D_1D_2}\left|D_2\sinh\sqrt{M}L_i-D_1\sinh\sqrt{M}L_i\right|
\end{aligned}
\end{equation}
Plugging this into the geodesic length \eqref{EEbr}, we obtain
\begin{eqnarray}
L_\gamma&\simeq& \log\left[\frac{2r^{(1)}r^{(2)}}{R^2}\frac{\cos\left(a|\Delta \tau|\right)-\cos\left(a|\Delta \varphi|\right)}{a^2}\right]\simeq\log\left[\frac{r^{(1)}r^{(2)}}{R^2}(|\Delta \varphi|^2-|\Delta \tau|^2)\right]\nn\\
&\simeq&2\log\left[\frac{\beta}{\pi z_\infty}\cosh\frac{\pi\Delta t}{\beta}\right]
\end{eqnarray}
In the late time regime, the interval differences equal
\begin{eqnarray}
&&|\Delta\tau|\simeq\pi-\frac{\sqrt{M}\epsilon}{D_1D_2}\left(D_2\sinh\sqrt{M}(t_-+t_\omega)-D_1\sinh\sqrt{M}(t_++t_\omega)\right)\\
&&|\Delta\varphi|\simeq\pi-\frac{\sqrt{M}\epsilon}{D_1D_2}\left(D_1\sinh\sqrt{M}L_i+D_2\sinh\sqrt{M}L_i\right)\,.
\end{eqnarray}
Since they are very close, we have
\begin{equation}
|\Delta\tau|\simeq |\Delta\tau|-\delta
\end{equation}
where
\begin{equation}
\delta\simeq\frac{\sqrt{M}\epsilon}{D_1D_2}\left[D_2(\sinh\sqrt{M}(t_- + t_\omega)-\sinh\sqrt{M}L_i)-D_1(\sinh\sqrt{M}(t_++t_\omega)+\sinh\sqrt{M}L_i)\right]
\end{equation}
This allows to write the geodesic length as
\begin{equation}
\begin{aligned}
  L_\gamma&\simeq \log\left[\frac{2r^{(1)}r^{(2)}}{R^2}\frac{\cos\left(a|\Delta \tau|\right)-\cos\left(a|\Delta \varphi|\right)}{a^2}\right]\simeq\log\left[\frac{2r^{(1)}r^{(2)}}{R^2}\frac{\sin\pi a}{a}\delta\right] \\
  &\simeq\log\left[\frac{\beta^2\frac{1}{2}\left(1+\cosh\frac{2\pi\Delta t}{\beta}\right)}{\pi^2z^2_\infty}\frac{2}{\sqrt{M}\epsilon}\frac{\sin\pi a}{a}\frac{\sinh\frac{\pi(t_-+t_\omega-L_i)}{\beta}\cosh\frac{\pi(L_i-t_+-t_\omega)}{\beta}}{\cosh\frac{\pi \Delta t}{\beta}}\right]\label{Lg2bd}
\end{aligned}
\end{equation}
where $\Delta t=t_--t_+$. These geodesics can now be used to compute the entanglement entropy of the union $S_{A\cup B}$.\\
In particular we will be interested in large $t_\omega>L_2 > L_1$ when $S_A=S_B=S_{\text{thermal}}$. In this case, there is a competition between the two geodesics connecting points in opposite boundaries and the geodesics connecting points in the same boundary giving rise to $2S_B$. The length of the new geodesics is
\begin{eqnarray}
L^1_\gamma&\simeq&\log\left[\left(\frac{\beta\cosh\frac{\pi\Delta t}{\beta}}{\pi z_\infty}\right)^2\frac{\beta}{\pi\epsilon}\frac{\sin\pi a}{a}\frac{\sinh\frac{\pi(t_-+t_\omega-L_1)}{\beta}\cosh\frac{\pi(L_1-t_+-t_\omega)}{\beta}}{\cosh\frac{\pi\Delta t}{\beta}}\right] \\
L^2_\gamma&\simeq&\log\left[\left(\frac{\beta\cosh\frac{\pi\Delta t}{\beta}}{\pi z_\infty}\right)^2\frac{\beta}{\pi\epsilon}\frac{\sin\pi a}{a}\frac{\sinh\frac{\pi(t_-+t_\omega-L_2)}{\beta}\cosh\frac{\pi(L_2-t_+-t_\omega)}{\beta}}{\cosh\frac{\pi\Delta t}{\beta}}\right]
\end{eqnarray}
where again $\Delta t=t_--t_+$.\\
Summarizing, the holographic entanglement entropy of the union of two intervals on the left and the right boundary is given by
\begin{equation}
S_{A\cup B}\simeq \frac{c}{6}\left(L^1_\gamma+L^2_\gamma\right),
\end{equation}
which matches with the CFT result \eqref{eq:saub}.

Finally from the above holographic results, we obtain the holographic mutual information $I_{A:B}=S_A+S_B-S_{A\cup B}$ and this again reproduces
the CFT result \eqref{eq:mutual_information_CFT} perfectly. As a consequence the scrambling time derived in the CFT \eqref{SCRtime} also holds as a result in gravity replacing $\alpha_\psi\to a$, as stressed below \eqref{EEbr}.

\subsection*{Acknowledgements}

We would like to thank Joan Camps, Tomoki Nosaka and K. Skenderis for discussions. PC is supported by  the Swedish Research Council (VR) grant 2013-4329. TT is supported by JSPS Grant-in-Aid for Scientific Research (B) No.25287058 and by JSPS Grant-in-Aid for Challenging Exploratory Research No.24654057. TT is also supported by World Premier International Research Center Initiative (WPI Initiative) from the Japan Ministry of Education, Culture, Sports, Science and Technology (MEXT). The work of JS and AS is supported by the Science and Technology Facilities Council (STFC) [grant number ST/L000458/1]. PC would like to thank the Galileo Galilei Institute for Theoretical Physics for the hospitality and the INFN for partial support during the completion of this work.

\begin{appendix}

\section{Twist operators in the TFD state}
\label{ap-A}

In this appendix we explain the proper ordering prescription of twist operator insertions when computing the two sided entanglement entropy $S_{A \cup B}$ (or mutual information $I_{A : B}$) in the thermofield double state.

\begin{figure}[!ht]
  \centering
  \includegraphics[natwidth=12cm, width=12cm]{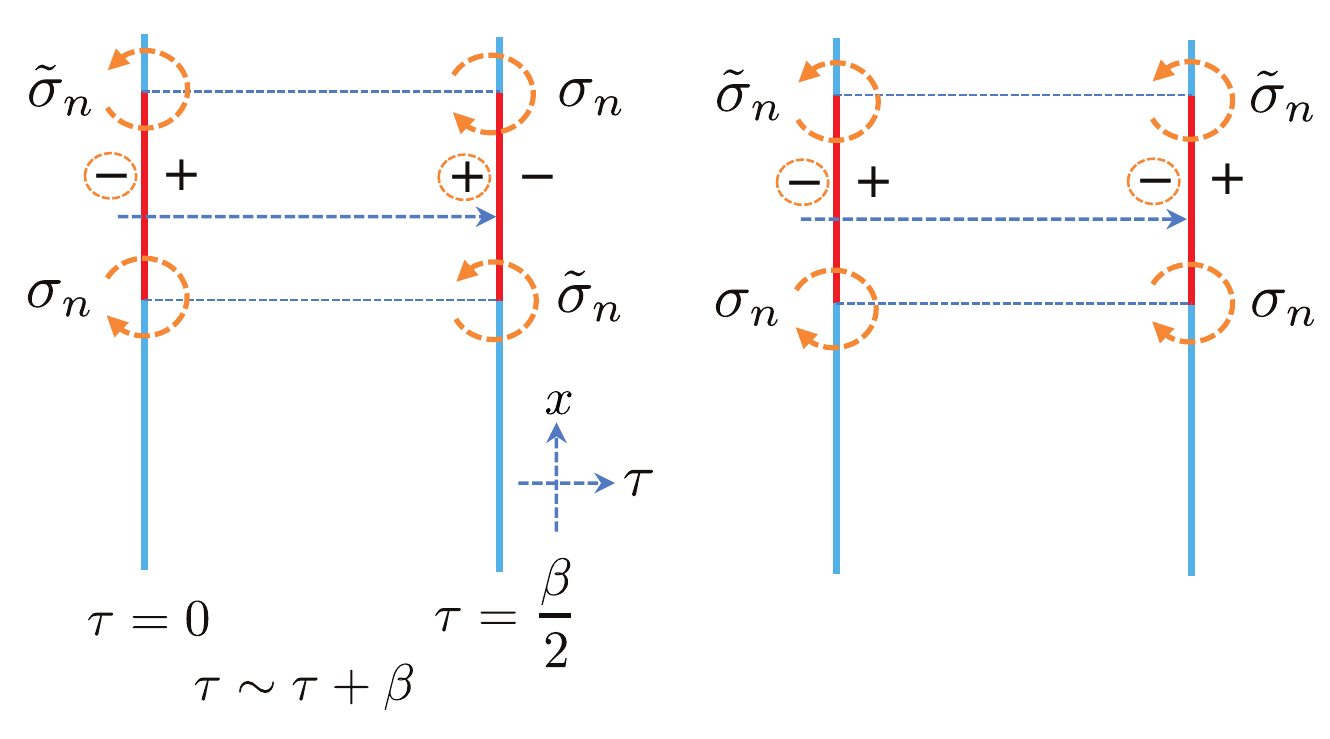}
  \caption{In the left picture (the second choice \eqref{order2}), one
returns to the starting point after going around two standard cylinders. In the
right picture (the first choice \eqref{order1}), this operation involves going around
$n$ standard cylinders.}\label{ns2}
\end{figure}

When computing $S_{A \cup B}$, with regions $A$ and $B$ belonging to opposite boundaries, the replica trick instructs us to insert
twist operators $\sigma_n , \tilde{\sigma}_n$ on each boundary. The order of these insertions is important, because different orderings can give rise to different replica geometries. There are two kinds of insertion orders. One, where both boundaries have the same order, such as
\begin{equation}
(\sigma_n, \tilde{\sigma}_n)_L, \ \ (\sigma_n, \tilde{\sigma}_n)_R\,.
\label{order1}
\end{equation}
In this case, going around the replica n-sheeted cylinder by
passing through the cuts between the twist operators on each boundary, one returns to the starting point after going around
$n$ cylinders (see the right picture in Fig.\ref{ns2}). There exists a second insertion order in which both boundaries have opposite twist operator orders, such as
\begin{equation}
(\sigma_n, \tilde{\sigma}_n)_L, \ \ (\tilde{\sigma}_n, \sigma_n)_R\,.
\label{order2}
\end{equation}
The same operation as above returns to the same point after going once around a standard cylinder (see the left picture in Fig.\ref{ns2}).

The question is what the right order prescription is when we consider the TFD state.
To answer this question, we provide two CFTs and one holographic bulk arguments.

It is convenient to remember the TFD state can be constructed by Euclidean time evolution
from the pure state $\sum_n |n\rangle_L |n\rangle_R$ with Euclidean time $\pm \beta/4$ for each boundary.
To compute the density matrix $\rho_{A \cup B}$, one glues partially by imposing boundary conditions to each sheet (half
cylinder) . Finally, to construct $\mathrm{Tr} \rho_{A \cup B}^n$, one glues the cylinders through their cuts on the cylinders, giving rise
to a partition function on the n-sheeted cylinder. If one goes around this n-sheeted cylinder through the cuts,
one returns to the starting point after going once around a standard cylinder because one should obtain the original state in the limit $\beta \to 0$.
Thus, the second insertion order (\ref{order2}) for twist operators is selected in this way.

\begin{figure}[!ht]
  \centering
  \includegraphics[natwidth=12cm, width=12cm]{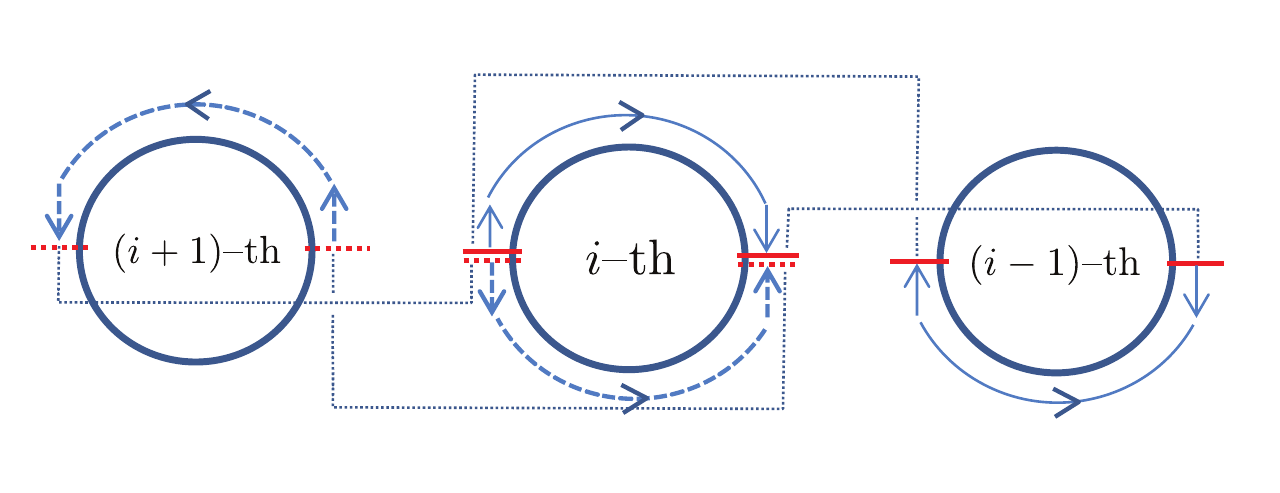}
  \caption{In the replica geometry which we obtain by gluing n
cylinders along the cuts, we can go around 2 standard cylinders, for example,
$i$-th and $(i+1)$-th cylinders (dotted line).}\label{ns1}
\end{figure}

A further argument to confirm this choice is as follows. The n-sheeted replica geometry is constructed by gluing single-sheets together along the cuts.
If one enters the cuts from the negative region, one emerges in the upper sheet, whereas if one enters from positive region, one ends in the lower sheet.
From this gluing condition, one obtains the n-sheeted replica geometry
corresponding to the second type of insertion order of twist operators (see Fig. \ref{ns1}).

Our final argument is holographic. In the bulk, one is instructed to consider geodesics connecting the different edges of the subsystems.
Twist operators are inserted in these edges. For geodesics connecting edges on the same boundary,  the inserted twist operators must have the same monodromy (or charge) properties, whereas for geodesics connecting edges in opposite boundaries, twist operators must have opposite monodromy (or charge). For example, if $\sigma_n$ is on an edge, $\tilde{\sigma}_n$ must appear at the other edge. This consistency condition chooses the same twist operator insertion order as in the CFT side.

\section{Details of the holographic model}
\label{ap-B}
Here we collect several useful formulas and conventions that we used in the part with holographic computations.\\

The relation between Kruskal and AdS-Schwarzschild coordinates can be obtained by referring both descriptions to the $\mathbb{R}^{2,2}$ where AdS$_3$ becomes the quadratic surface
\begin{equation}
 -X_0^2-X_1^2 + X_2^2+X_3^2=-R^2
 \end{equation}
in which we have
\begin{equation}
\begin{aligned}
  \pm\frac{R\sqrt{1-Mz^2}}{\sqrt{M}z}\sinh\left(\sqrt{M}t_{\mp}\right) &=X_0 = R\frac{u+v}{1+uv}=R\frac{2T}{1+T^2-X^2}\,, \\
  \frac{R}{\sqrt{M}z}\cosh\left(\sqrt{M}\theta\right) &=X_1 = R\frac{1-uv}{1+uv}\cosh\phi=R\frac{1-T^2+X^2}{1+T^2-X^2}\,\cosh\phi\,, \\
  \frac{R}{\sqrt{M}z}\sinh\left(\sqrt{M}\theta\right) &= X_2 = R\frac{1-uv}{1+uv}\sinh\phi=R\frac{1-T^2+X^2}{1+T^2-X^2}\,\sinh\phi\,, \\
  \pm\frac{R\sqrt{1-Mz^2}}{\sqrt{M}z}\cosh\left(\sqrt{M}t_{\mp}\right) &= X_3 = R\frac{u-v}{1+uv}=-R\frac{2X}{1+T^2-X^2}\,.
\end{aligned}
\label{MapN}
\end{equation}
This formulas fix our conventions for the appropriate signs on the gravity side.\\

In order to derive lengths of the geodesic in the back-reacted metric we only need to know the asymptotic form of the map from the left and the right wedge to $AdS_3$ in global coordinates. This is just the map that takes the trajectory of our massive point particle in Kruskal coordinates to the $r=0$ particle in global $AdS_3$. In the left and right exteriors the map becomes
\begin{equation}
\begin{aligned}
\sqrt{R^2+r^2}\sin\tau&=\pm\frac{R\sqrt{1-Mz^2_\infty}}{\sqrt{M}z_\infty} \sinh\left(\sqrt{M}(t_\mp+t_\omega)\right)\,,\\
\sqrt{R^2+r^2}\cos\tau&=\frac{R}{M\epsilon z_\infty}\left(\cosh\left(\sqrt{M}\theta\right)\mp\sqrt{1-M\epsilon^2}\sqrt{1-Mz^2_\infty}\cosh\left(\sqrt{M}(t_\mp+t_\omega)\right)\right)\,,\\
r\sin\varphi&=\frac{R}{\sqrt{M}z_\infty}\sinh\left(\sqrt{M}\theta\right)\,,\\
r\cos\varphi&=\frac{R}{M\epsilon z_\infty}\left(\pm\sqrt{1-Mz^2_\infty}\cosh\left(\sqrt{M}(t_\mp+t_\omega)\right)-\sqrt{1-M\epsilon^2}\cosh\left(\sqrt{M}\theta\right)\right)\,.
\end{aligned}
\label{MapN1}
\end{equation}
The points in the main text are extracted to the first order in $\epsilon$.\\
Note that the map from the right wedge also depends on $\epsilon$ as well as $t_\omega$.

\section{Two sided 2-pt functions}
\label{ap-C}

Given the bound \eqref{eq:bound} that the mutual information provides on the amount of correlation, it is natural to study the two sided two-point correlation function in the TFD after a local perturbation is turned on in one of the boundaries. This is the calculation described in this appendix.

Given a local primary probe operator $O_h(x,\bar{x})$ of conformal dimension $h$ and a perturbation described by a different primary $O_{h_w}$, the appropriate normalised 4-point two sided correlation is
\begin{equation}
C_4=\frac{\langle O_{h_w}(x_1,\bar{x}_1)O_{h}(x_2,\bar{x}_2)O_{h}(x_3,\bar{x}_3)O_{h_w}(x_4,\bar{x}_4)\rangle}{\langle O_{h_w}(x_1,\bar{x}_1)O_{h_w}(x_4,\bar{x}_4)\rangle}
\end{equation}
where the insertion points are
\begin{eqnarray}
&&x_1=-i\epsilon\qquad\qquad x_2=L_1-t_--t_\omega,\qquad x_3=L_2-t_+-t_\omega+i\frac{\beta}{2}\qquad x_4=i\epsilon\nn\\
&&\bar{x}_1=i\epsilon\qquad \bar{x}_2=L_1+t_-+t_\omega,\qquad \bar{x}_3=L_2+t_++t_\omega-i\frac{\beta}{2}\qquad \bar{x}_4=-i\epsilon
\end{eqnarray}
Assuming $h/c\ll 1$ and keeping $h_w/c$ fixed, the numerator is again a 4-point function involving two heavy and two light operators. Using the large central charge results of \cite{Fitzpatrick:2014vua} we can write the correlator as
\begin{equation}
C_4=\left|\frac{\beta}{\pi z_\infty}\sinh\frac{\pi x_{23}}{\beta}\right|^{-4h}\left|1-z\right|^{4h}G(z,\bar{z})
\end{equation}
with
\begin{equation}
G(z,\bar{z})\simeq \left(\frac{z^{\frac{1-\alpha}{2}}(1-z^{\alpha})\bar{z}^{\frac{1-\alpha}{2}}(1-\bar{z}^{\alpha})}{\alpha^2}\right)^{-2h},\qquad \alpha=\sqrt{1-\frac{24h_\omega}{c}}\,.
\end{equation}
In the limit of small $\epsilon/\beta$, the cross-ratios reduce to
\begin{eqnarray}
z\simeq 1+\frac{2\pi i \epsilon}{\beta}\frac{\cosh\frac{\pi(\Delta L+\Delta t)}{\beta}}{\sinh\frac{\pi(L_1-t_--t_\omega)}{\beta}\cosh\frac{\pi(L_2-t_+-t_\omega)}{\beta}}\\
\bar{z}\simeq 1-\frac{2\pi i \epsilon}{\beta}\frac{\cosh\frac{\pi(\Delta L+\Delta t)}{\beta}}{\sinh\frac{\pi(L_1+t_-+t_\omega)}{\beta}\cosh\frac{\pi(L_2+t_++t_\omega)}{\beta}}
\end{eqnarray}
where $\Delta L=L_2-L_1$ and $\Delta t=t_--t_+$. As in our main text discussions, the key observation is that the imaginary part of $z$ depends on the sign of
$L_1-t_--t_\omega$. For $L_1>t_-+t_\omega$, we have $(z,\bar{z})\sim (1,1)$ and
\begin{equation}
C_4\simeq\left(\frac{\beta}{\pi z_\infty}\sqrt{\frac{1}{2}\left(\cosh\frac{2\pi \Delta L}{\beta}+\cosh\frac{2\pi\Delta t}{\beta}\right)}\right)^{-4h}
\end{equation}
whereas for $L_1<t_-+t_\omega$, $(z,\bar{z})\sim (e^{2\pi i},1)$ and
\begin{equation}
C_4\simeq\left(\frac{\beta}{\pi z_\infty}\sqrt{\frac{1}{2}\left(\cosh\frac{2\pi \Delta L}{\beta}+\cosh\frac{2\pi\Delta t}{\beta}\right)}\right)^{-4h}\left(\frac{\beta}{\pi\epsilon}\frac{\sin(\pi\alpha)}{\alpha}\frac{\sinh\frac{\pi(t_-+t_\omega-L_1)}{\beta}\cosh\frac{\pi(L_2-t_+-t_\omega)}{\beta}}{\cosh\frac{\pi(\Delta L+\Delta t)}{\beta}}\right)^{-2h}\label{2ptL}
\end{equation}
It can be checked this result precisely matches the gravity computation where the two-point function is given by the length of a geodesic \eqref{Lg2bd} between two-boundaries in our back-reacted metric.

One can read off the scrambling time scale from the two-point correlators. For example, setting $L_1=L_2=0$ and $t_-=t_+=0$, the correlation $\eqref{2ptL}$ for large $t_\omega$ is given by
\begin{equation}
C_4\simeq \left(\frac{\beta}{2\pi z_\infty}\right)^{-4h}\exp\left[-\frac{4\pi h}{\beta}\left(t_w+\frac{\beta}{2\pi}\log\left(\frac{\beta}{\pi\epsilon}\frac{\sin(\pi\alpha)}{\alpha}\right)\right)\right]\,.
\end{equation}
This reproduces the dependence on the perturbation for the scrambling time derived in the mutual information analysis.

\end{appendix}



\begin{thebibliography}{}
\bibitem{cirac}
M.~Wolf, F.~Verstraete, M.~Hastings, and J.I.~ Cirac. "Area laws in quantum systems: mutual information and correlations," Physical Review Letters {\bf 100}, no. 7 (2008): 070502.

\bibitem{Witten:1998zw}
  E.~Witten,
  ``Anti-de Sitter space, thermal phase transition, and confinement in gauge theories,''
  Adv.\ Theor.\ Math.\ Phys.\  {\bf 2} (1998) 505
  [hep-th/9803131].

  \bibitem{eternal}
  J.~M.~Maldacena,
  ``Eternal black holes in anti-de Sitter,''  JHEP {\bf 0304} (2003) 021  [hep-th/0106112].

\bibitem{Hayden:2007cs}
 P.~Hayden and J.~Preskill,
  ``Black holes as mirrors: Quantum information in random subsystems,''  JHEP {\bf 0709} (2007) 120  [arXiv:0708.4025 [hep-th]].  

\bibitem{Sekino:2008he}
  Y.~Sekino and L.~Susskind,
  ``Fast Scramblers,''
  JHEP {\bf 0810} (2008) 065
  [arXiv:0808.2096 [hep-th]].

  \bibitem{Shenker:2013pqa}
  S.~H.~Shenker and D.~Stanford,
  ``Black holes and the butterfly effect,''
  JHEP {\bf 1403} (2014) 067
  [arXiv:1306.0622 [hep-th]].

\bibitem{Shenker:2013yza}
  S.~H.~Shenker and D.~Stanford,
  ``Multiple Shocks,''
  JHEP {\bf 1412} (2014) 046
  [arXiv:1312.3296 [hep-th]].

\bibitem{Shenker:2014cwa}
  S.~H.~Shenker and D.~Stanford,
  ``Stringy effects in scrambling,''
  arXiv:1412.6087 [hep-th].

  \bibitem{BTZ}
  M.~Banados, C.~Teitelboim and J.~Zanelli,
  ``The Black hole in three-dimensional space-time,''
  Phys.\ Rev.\ Lett.\  {\bf 69} (1992) 1849
  [hep-th/9204099].

\bibitem{Fitzpatrick:2014vua}
  A.~L.~Fitzpatrick, J.~Kaplan and M.~T.~Walters,
  ``Universality of Long-Distance AdS Physics from the CFT Bootstrap,''
  JHEP {\bf 1408} (2014) 145
  [arXiv:1403.6829 [hep-th]].

\bibitem{Fitzpatrick:2015zha}
  A.~L.~Fitzpatrick, J.~Kaplan and M.~T.~Walters,
  ``Virasoro Conformal Blocks and Thermality from Classical Background Fields,''
  arXiv:1501.05315 [hep-th].

\bibitem{Hartman:2013mia}
  T.~Hartman,
  ``Entanglement Entropy at Large Central Charge,''  arXiv:1303.6955 [hep-th].

\bibitem{Hijano:2015rla}
  E.~Hijano, P.~Kraus and R.~Snively,
  ``Worldline approach to semi-classical conformal blocks,''
  arXiv:1501.02260 [hep-th].


\bibitem{cal}
P.~Calabrese and J.~L.~Cardy,
  ``Entanglement and correlation functions following a local quench: a conformal field theory approach,''
  J.\ Stat.\ Mech.\  {\bf 0710}  P10004, arXiv:0708.3750.


\bibitem{Nozaki:2014hna}
  M.~Nozaki, T.~Numasawa and T.~Takayanagi,
  ``Quantum Entanglement of Local Operators in Conformal Field Theories,''  Phys.\ Rev.\ Lett.\  {\bf 112} (2014) 111602  [arXiv:1401.0539 [hep-th]].  

\bibitem{He:2014mwa}
  S.~He, T.~Numasawa, T.~Takayanagi and K.~Watanabe,
  ``Quantum dimension as entanglement entropy in two dimensional conformal field theories,''  Phys.\ Rev.\ D {\bf 90} (2014) 4,  041701  [arXiv:1403.0702 [hep-th]].  

\bibitem{Nozaki:2014uaa}
  M.~Nozaki,
  ``Notes on Quantum Entanglement of Local Operators,''  JHEP {\bf 1410} (2014) 147  [arXiv:1405.5875 [hep-th]].  

  \bibitem{NNT}
  M.~Nozaki, T.~Numasawa and T.~Takayanagi,
  ``Holographic Local Quenches and Entanglement Density,''  JHEP {\bf 1305} (2013) 080  [arXiv:1302.5703 [hep-th]].

  \bibitem{Caputa:2014vaa}
  P.~Caputa, M.~Nozaki and T.~Takayanagi,
  ``Entanglement of local operators in large-N conformal field theories,''
  PTEP {\bf 2014}, no. 9, 093B06 (2014)
  [arXiv:1405.5946 [hep-th]].

 \bibitem{Asplund:2014coa}
  C.~T.~Asplund, A.~Bernamonti, F.~Galli and T.~Hartman,
  ``Holographic Entanglement Entropy from 2d CFT: Heavy States and Local Quenches,''
  arXiv:1410.1392 [hep-th].

\bibitem{Giusto:2014aba}
  S.~Giusto and R.~Russo,
  ``Entanglement Entropy and D1-D5 geometries,''  Phys.\ Rev.\ D {\bf 90} (2014) 6,  066004  [arXiv:1405.6185 [hep-th]].  

\bibitem{deBoer:2014sna}
  J.~de Boer, A.~Castro, E.~Hijano, J.~I.~Jottar and P.~Kraus,
  ``Higher Spin Entanglement and WN Conformal Blocks,''  arXiv:1412.7520 [hep-th].  

\bibitem{Guo:2015uwa}
  W.~Z.~Guo and S.~He,
  ``R\'enyi entropy of locally excited states with thermal and boundary effect in 2D CFTs,''  arXiv:1501.00757 [hep-th].  

\bibitem{Caputa:2014eta}
  P.~Caputa, J.~Sim\'on, A.~\v{S}tikonas and T.~Takayanagi,
  ``Quantum Entanglement of Localized Excited States at Finite Temperature,''
  JHEP {\bf 1501} (2015) 102
  [arXiv:1410.2287 [hep-th]].


\bibitem{Ugajin:2013xxa}
  T.~Ugajin,
  ``Two dimensional quantum quenches and holography,''  arXiv:1311.2562 [hep-th].  

\bibitem{Asplund:2013zba}
  C.~T.~Asplund and A.~Bernamonti,
  ``Mutual information after a local quench in conformal field theory,''  Phys.\ Rev.\ D {\bf 89} (2014) 6,  066015  [arXiv:1311.4173 [hep-th]].  

  \bibitem{Roberts:2014isa}
  D.~A.~Roberts, D.~Stanford and L.~Susskind,
  ``Localized shocks,''
  arXiv:1409.8180 [hep-th].



\bibitem{RTorig}
  S.~Ryu and T.~Takayanagi,
  ``Holographic derivation of entanglement entropy from AdS/CFT,''
  Phys.\ Rev.\ Lett.\  {\bf 96} (2006) 181602

\bibitem{Hubeny:2007xt}
  V.~E.~Hubeny, M.~Rangamani and T.~Takayanagi,
  ``A Covariant holographic entanglement entropy proposal,''
  JHEP {\bf 0707}, 062 (2007)
  [arXiv:0705.0016 [hep-th]].


  \bibitem{Roberts:2014ifa}
  D.~A.~Roberts and D.~Stanford,
  ``Two-dimensional conformal field theory and the butterfly effect,''
  arXiv:1412.5123 [hep-th].


  \bibitem{Maldacena:2015waa}
  J.~Maldacena, S.~H.~Shenker and D.~Stanford,
  ``A bound on chaos,''
  arXiv:1503.01409 [hep-th].


\bibitem{CC}
 P.~Calabrese and J.~L.~Cardy,
  ``Entanglement entropy and quantum field theory,''
  J.\ Stat.\ Mech.\  {\bf 0406}, P06002 (2004)
  [hep-th/0405152].

  \bibitem{Cardy:2007mb}
  J.~L.~Cardy, O.~A.~Castro-Alvaredo and B.~Doyon,
  ``Form factors of branch-point twist fields in quantum integrable models and entanglement entropy,''
  J.\ Statist.\ Phys.\  {\bf 130}, 129 (2008)
  [arXiv:0706.3384 [hep-th]].

\bibitem{cag}
P.~Calabrese and J.~L.~Cardy,
  ``Evolution of Entanglement Entropy in One-Dimensional Systems,''
  J.\ Stat.\ Mech.\  {\bf 04} (2005) P04010, cond-mat/0503393.

\bibitem{Mor}
  I.~A.~Morrison and M.~M.~Roberts,
  ``Mutual information between thermo-field doubles and disconnected holographic boundaries,''
  arXiv:1211.2887 [hep-th].

\bibitem{HaMa}
  T.~Hartman and J.~Maldacena,
  ``Time Evolution of Entanglement Entropy from Black Hole Interiors,''  JHEP {\bf 1305} (2013) 014  [arXiv:1303.1080 [hep-th]].

\bibitem{Caputa:2013eka}
  P.~Caputa, G.~Mandal and R.~Sinha,
  ``Dynamical entanglement entropy with angular momentum and U(1) charge,''
  JHEP {\bf 1311}, 052 (2013)
  [arXiv:1306.4974 [hep-th]].


\bibitem{Calabrese:2010he}
  P.~Calabrese, J.~Cardy and E.~Tonni,
  ``Entanglement entropy of two disjoint intervals in conformal field theory II,''
  J.\ Stat.\ Mech.\  {\bf 1101} (2011) P01021
  [arXiv:1011.5482 [hep-th]].

\bibitem{Perlmutter:2013paa}
  E.~Perlmutter,
  ``Comments on Renyi entropy in AdS$_3$/CFT$_2$,''
  JHEP {\bf 1405} (2014) 052
  [arXiv:1312.5740 [hep-th]].

\bibitem{Horowitz:1999gf}
  G.~T.~Horowitz and N.~Itzhaki,
  ``Black holes, shock waves, and causality in the AdS / CFT correspondence,''
  JHEP {\bf 9902}, 010 (1999)
  [hep-th/9901012].

\end{thebibliography}
\end{document}